\documentclass[twocolumn,floatfix,showpacs,prd,aps,superscriptaddress]{revtex4-2}
\usepackage{amsmath,amsfonts}
\usepackage{epsfig,subfigure}
\usepackage{ifpdf}
\usepackage{url}
\usepackage[colorlinks=true, citecolor=cyan, linkcolor=cyan, urlcolor= cyan]{hyperref} \usepackage{amsmath}
\DeclareMathOperator{\arctantwo}{arctan2}
\usepackage{physics}
\usepackage{color}
\usepackage{natbib}
\usepackage{multirow}
\usepackage{graphicx}
\usepackage{dcolumn}
\usepackage{bm}
\usepackage{rotating}
\usepackage{float}
\usepackage{xcolor}
\usepackage[utf8]{inputenc}
\usepackage{mathtools}
\usepackage{academicons}
\definecolor{orcidlogocol}{HTML}{A6CE39}

\newcommand{\apjl}{The Astrophysical Journal Letters}
\newcommand{\araa}{Annual Review of Astronomy and Astrophysics}
\newcommand{\NA}{---}
\begin{document}


\title{Testing the Black Hole No-hair Theorem with Galactic Center Stellar Orbits
}

\def\addUWM{Center for Gravitation Cosmology and Astrophysics, University of Wisconsin-Milwaukee, Milwaukee, WI 53201, USA}

\def\addRIT{Center for Computational Relativity and Gravitation, Rochester Institute of Technology, Rochester, NY 14623, USA}

\def\addCardiff{School of Physics and Astronomy, Cardiff University, Cardiff CF24 3AA, United Kingdom}

\author{Hong Qi}
\email{hong.qi@ligo.org}
\affiliation{\addCardiff}
\affiliation{\addUWM}

\author{Richard O'Shaughnessy}
\affiliation{\addRIT}

\author{Patrick Brady}
\affiliation{\addUWM}

\date{\today}

\begin{abstract}
Theoretical investigations have provided proof-of-principle calculations suggesting measurements of stellar or pulsar orbits near the Galactic Center could strongly constrain the properties of the Galactic Center black hole, local matter, and even the theory of gravity itself.  As in previous studies, we use a Markov chain Monte Carlo to quantify what properties of the Galactic Center environment  measurements can constrain.  In this work, however, we also develop an analytic model (Fisher matrix) to understand what parameters are well-constrained and why.  Using both tools, we conclude that existing astrometric measurements cannot constrain the spin of the Galactic Center black hole.  Extrapolating to the precision and cadence of future experiments, we anticipate that the black hole spin can be measured with the known star S2. Our calculations show that we can measure the dimensionless black hole spin to a precision of $\sim$0.1 
with weekly measurements of the orbit of S2 for 40 years using the GRAVITY telescope's best resolution at the Galactic Center, i.e., an angular resolution of $10\ \mu$arcsecond and a radial velocity resolution of 500 m/s. An analytic expression is derived for the measurement uncertainty of the black hole spin using Fisher matrix in terms of observation strategy, star's orbital parameters, and instrument resolution. From it we conclude that highly eccentric orbits can provide better constraints on the spin, and that an orbit with a higher eccentricity is more favorable even when the orbital period is longer. 
We also apply it to S62, S4711, and S4714 to show whether they can constrain the black hole spin sooner than S2. If in addition future measurements include discovery of a new, tighter stellar orbit, then future data could conceivably enable tests of strong field gravity, by directly measuring the black hole quadrupole moment. Our simulations show that with a stellar orbit similar to that of S2 but at one fifth the distance to the Galactic Center and GRAVITY's resolution limits on the Galactic Center, we can start to test the no-hair theorem with 20 years of weekly orbital measurements. 
\end{abstract}

\maketitle

\section{Introduction}

The supermassive black hole at the center of our galaxy provides unique opportunities to investigate dynamics near a strongly-gravitating source \citep{2011JPhCS.283a2030P,2019BAAS...51c.530D}. 
Radio telescopes have imaged the immediate vicinity of the black hole \citep{2000ApJ...528L..13F,Akiyama:2019cqa}, allowing
direct constraints on the strong gravitational field regime near the black hole via imaging accretion flows \citep{Psaltis:2018xkc,Reynolds:2019uxi,Dokuchaev:2019bbf,Bambi:2019tjh,2020AAS...23536915D,Psaltis:2020lvx}.
Stellar motions also constrain the number and orbits of nearby perturbers \citep{2011ApJ...735...57B}. 
At present, however, the best opportunities to constrain the Galactic Center come from long-term monitoring of known
stars \citep{2005ApJ...620..744G,2008ApJ...689.1044G,2009ApJ...692.1075G,2012Sci...338...84M,2017ApJ...837...30G}.  
These measurements can also identify effects from the strong gravitational field
\citep{2006ApJ...639L..21Z,2007ApJ...654L..83Z,2020ApJ...888L...8N} 
and the properties of the supermassive black hole
\citep{2005ApJ...620..744G,2005ApJ...622..878W,2008ApJ...674L..25W,2008ApJ...689.1044G,2009ApJ...692.1075G,2010PhRvD..81f2002M,will2016,2019BAAS...51c.530D,Abuter:2020dou,Fragione:2020khu}.  Even stronger constraints would be possible with a
well-timed pulsar orbiting the Galactic Center
\citep{2012ApJ...747....1L,Zhang:2017qbb,Wex:2012au,2012ApJ...753..108W,Psaltis:2015uza}, at separations comparable to a
recently-discovered object  \citep{2013ApJ...775L..34R}. 
High precision inference from stellar orbits ideally should account for many nearby perturbers, including the local
stellar density of visible stars \citep{2013ApJ...764..155L} and compact objects \citep{2010PhRvD..81f2002M}. 

Motivated by recent discoveries of new stars in close orbits around the Galactic Center \citep{2020ApJ...899...50P}, we assess how well existing and future measurements of stellar orbits \citep{2019BAAS...51c.530D} can constrain the black hole properties: its mass and particularly its spin.   Specifically, we wrote a Markov chain Monte Carlo (MCMC) code and use it to compare real and synthetic astrometric and radial velocity data with models for the stellar orbits and black hole mass,  accounting for differences in reference frame between different observational campaigns. Unlike previous investigations, our model includes leading order post Newtonian corrections to the orbit from the black hole's mass, spin, and quadrupole moment, as well as the impact of unknown non-quadrupole internal and exterior potentials.    Our goal is to determine whether, despite the extremely low orbital velocity $v/c\simeq 0.02$, future measurements can significantly constrain strong-field features of the Galactic Center black hole.   
We compare our MCMC results against a detailed Fisher matrix analysis, both to validate our results and allow the reader to easily extrapolate to future measurement scenarios.

This paper is organized as follows.  In Section \ref{sec:Methods} we review the observations of stellar orbits near the Galactic Center; review a simplified model for stellar dynamics near supermassive black holes (justified at length in Appendix \ref{ap:EOMs}); introduce simplified and realistic models for the process of measuring stellar orbits, including errors. In Section  \ref{sec:Fisher} we describe two techniques to assess how
well measurements can constrain properties of stellar orbits and the supermassive black hole.  
The first is a simplified, approximate Fisher matrix.  
The second method uses detailed
Markov chain Monte Carlo simulations of synthetic data to determine how well different parameters can be measured and
why.  After validating our procedure using analytically tractable toy models with a handful of parameters, we perform
full-scale simulations in Section \ref{sec:hypothesestests} to test several hypotheses including no-hair theorem. 
Using plausible choices of parameters and future achievable measurement accuracy, we discuss how the black hole spin and
quadrupole moment can be constrained with the known star S2 of an orbital period of about 16 years at about $5 \text{mpc}$ distance from the black hole and future discoverable closer stars with orbital periods as small as 1-2 years at about $1\ \text{mpc}$ separation \citep{Graham:2019mis}. We also discuss how these constraints can be affected by an intermediate-mass black hole (IMBH) and a cluster of other stars in the Galactic Center. 
In Section \ref{sec:conclusion} we summarize the conclusions we draw from the studies.
Throughout the paper we adopt the units where $G = c = 1$.

\section{Statement of the problem}
\label{sec:Methods}
\subsection{Existing Observations}

There are observations of stellar orbits within 1 arcsecond of the Galactic Center in infrared \citep{2008ApJ...689.1044G,2009ApJ...692.1075G,2020ApJ...899...50P,2017ApJ...837...30G}. In this paper, we are analyzing two sets of long-duration observations reported in Ghez et al. \citep{2008ApJ...689.1044G} and Gillessen et al. \citep{2009ApJ...692.1075G}. The motions of stars in the immediate vicinity of Sgr A* have been observed in infrared bands by NTT/VLT since 1992 and by Keck telescope since 1995. The two data sets we use are the Keck data from 1995 to 2007 and the VLT data from 1992 to 2009. Massive young stars are found closely orbiting the black hole at the center of our Milky Way. The locations of the stars, i.e., the astrometric positions, right ascensions (RA) and declinations (DEC) are recorded at different epochs. 
Therefore, the relative positions of stars to the radio source Sgr A* , i.e., the offsets of RA and DEC are also measured. The radial velocities, i.e., the line of sight components of the velocities relative the the observers, of each star at different epochs are also measured. In this work, we use the stellar orbit of star S2, because it is monitored for the longest time, its orbit is only 16 years, and more importantly its eccentricity is high among the few closest orbits that have been monitored frequently for over a decade. The high eccentricity makes the star get deeper in the gravitational potential of the black hole and thus can provide more physics. We show in \ref{sec:measureblackholespin} with concrete simulations why the orbit of S2 provides better constraints than that of S102/S55 (S102 is short for S0-102 \citep{Meyer:2012hn} which was a previous name of S55) even though the latter has a smaller orbital period (12 years) and even if they were observed the same way. There have been more recent observations and measurements of the S2 stellar orbit \citep{2017PhRvL.118u1101H,2019ApJ...873....9J,2019Sci...365..664D} as we prepared our paper, but the added data do not affect our conclusions.

\subsection{Simplified models of stellar orbits}

The approximations involved in deriving and justifying our equations of motion are provided in Appendix \ref{ap:EOMs}.  
Neglecting the black hole's recoil or the effect of ambient material, each star's position $\mathbf{x}$ evolves according to leading-order post-Newtonian equations of motion  \citep{2008ApJ...674L..25W, book-merritt-GalacticCenter, book-Will-TestingGR}
\begin{eqnarray}
{\bf a} = -\frac{M {\bf x}}{r^3}+\frac{M {\bf x}}{r^3}(4\frac{M}{r}-v^2)+4\frac{M \dot r}{r^2}{\bf v} \nonumber\\
-\frac{2J}{r^3}[2{\bf v}\times {\bf\hat{J}} - 3 \dot r {\bf \hat n} \times {\bf\hat{J}} - 3 {\bf \hat n} ({\bf L \cdot \hat J})/r] \nonumber\\
+\frac{3}{2}\frac{Q_2}{r^4}[5 {\bf \hat n ( \hat n \cdot \hat J})^2 - 2({\bf \hat n \cdot \hat J)\hat J - \hat n}],
\label{eq:EOM:SingleBody}
\end{eqnarray}
where ${\bf x}, {\bf v}=\partial_t {\bf x},{\bf a}=\partial_t^2 {\bf x}$ are the harmonic coordinate position, velocity, and acceleration of the star, $r=|{\bf x}|$ is the coordinate distance of the star from the black hole, ${\bf \hat n} = {\bf x}/r$ is a unit vector pointing towards
the star, ${\bf L}={\bf x} \times {\bf v}$ is the orbital angular momentum, $M,{\bf J},{ Q_2=-J^2/M}$ are the mass, spin angular momentum, and quadrupole moment of the black hole, and the hat over a quantity denotes its unit vector, such as ${\bf\hat J}={\bf{J}}/J$.  
Each star evolves according to a post-Newtonian Hamiltonian in \citep{FlanaganTichy}. 

\begin{figure}
\includegraphics[width=\columnwidth]{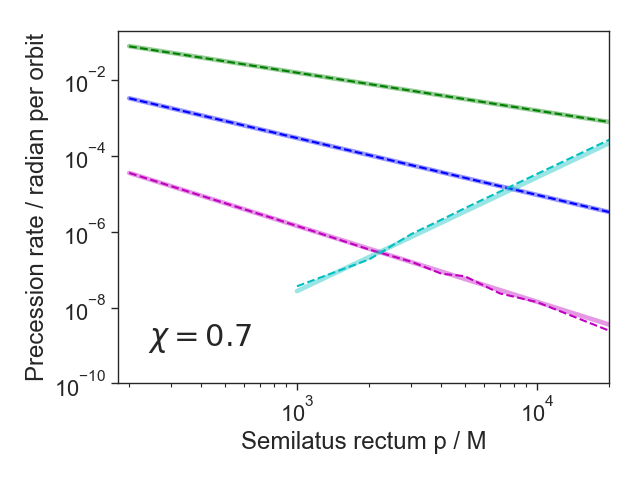}
\caption{\label{fig:PrecessionRates}Relative magnitudes of characteristic precession rates as a function of semilatus rectum for a stellar orbit around the supermassive black hole due to different effects. Solid curves show analytic results; dotted curves are derived from our time-domain evolution code, as validations. The solid green, blue, an purple curves show $A_S$, $A_J$, and $A_Q$ that are derived in \citep{2008ApJ...674L..25W} and implicitly defined in Eqs.~\eqref{eq:AS}, \eqref{eq:AJ} and \eqref{eq:AQ}.  The cyan curves show the influence of an external quadrupolar potential from a cluster of ambient stars of mass $200\ M_\odot$ at a distance of $3\times 10^4$ times the black hole mass $M$. }
\end{figure}

For the proof-of-concept analytic calculations, we separate timescales by \emph{orbit-averaging} rather than work with the full Hamiltonian, following standard practice in celestial mechanics.  For analytic simplicity, we will furthermore treat all
perturbations at leading order,  therefore performing an orbit average using a  Newtonian orbit; for example, at leading
order an equatorial  orbit has the form $r(t)=p/(1+e\cos \Phi(t))$, where $p=a(1-e^2)$ is a semilatus rectum, $a$ is the semimajor axis, $e$ is the eccentricity of the orbit, and $\Phi(t)$ is the orbital phase in terms of time $t$.  Using standard methods of celestial mechanics
\citep{2008ApJ...674L..25W,2011CQGra..28v5029S}, we find the secular equations of motion for the orbit average
($\left<X\right>$) of each star's Newtonian orbital angular momentum ${\bf{L}}_N \equiv \mu {\bf x} \times {\bf v}$  and Newtonian Runge-Lenz vector ${\bf A}_N \equiv \mu^2[{\bf v}\times ({\bf x}\times {\bf v})- G M{\bf{\hat{n}}}]$:
\begin{align}
\label{eq:OrbitAverage:SecularEOMForConservedQuantities}
&\partial_t \left<{\bf L}_N \right> =  \vec{\Omega} \times\left< {\bf L}_N \right>\\
&\partial_t \left<{\bf A}_N \right> =  \vec{\Omega} \times \left<{\bf A}_N \right>\\
&\vec\Omega = \vec\Omega_S +\vec\Omega_J + \vec\Omega_Q \\
\label{eq:AS}
&\vec\Omega_S = {\bf\hat{L}}_N \frac{A_S} {P} = {\bf\hat{L}}_N \frac{3 }{p  (a/M)^{\frac{3}{2}} } \\
\label{eq:AJ}
&\vec\Omega_J = [\hat{\bf J} -3{\bf\hat{L}}({\bf\hat{L}}\cdot{\bf\hat{J}})] \frac{A_J}{ P}
= [{\bf\hat{J}} - 3{\bf\hat{L}}({\bf\hat{L}}\cdot {\bf\hat{J}})] \frac{2 J/M}{(Mp^3)^{\frac{1}{2}} (\frac{a}{M})^{\frac{3}{2}}} \\
\label{eq:AQ}
&\vec\Omega_Q = -({\bf\hat{J}}( {\bf\hat{J}} \cdot {\bf\hat{L}}) + \frac{1}{2}{\bf\hat{L}}(1-3 ({\bf\hat{L}}\cdot {\bf\hat{J}})^2) \frac{ A_Q}{P} \\
&A_Q = \frac{3}{2} \frac{Q_2}{ p^2 (a/M)^{3/2}},
\end{align}
where the expressions $\vec\Omega$, $\vec\Omega_S$, $\vec\Omega_J$, and $\vec\Omega_Q$ are the orbital precession, $P$ is the orbital period, the expressions $A_S$, $A_J$, and $A_Q$ derived in \citep{2008ApJ...674L..25W} are implicitly defined here; see
also \citep{2011PhRvD..84l4001I}. 
The factors $A_S$, $A_J$, and $A_Q$ are shown in Figure \ref{fig:PrecessionRates}.
Note that the dimensionless spin $\chi=J/M^2$ is used and it is always less than one. These orbit-averaged precession equations imply a straightforward procedure for the linear perturbation
due to $\Omega$, starting from a Newtonian solution $\vec{r}_o(t)$:
\begin{eqnarray}
\vec{r}(t) \simeq R(t) \vec{r}_o(t),
\end{eqnarray}
where $R(t)$ is the rotation generated by the orbit-averaged $\vec{\Omega}$.  Specifically, again working to first order in the orbit-averaged perturbations, the secular rotation $R(t)$ on short timescales is determined by the
generators ${\cal L}_\alpha$ of rotations:
\begin{align}
R(t)  &\simeq {\bf 1} - i t {\cal L}_\alpha \Omega^\alpha \\
\label{eq:OrbitAveragedSolution:SecularRotation}
\vec{r}(t)& \simeq r_o(t)  - i t  \Omega^\alpha{\cal L}_\alpha r_o(t).
\end{align}

\subsection{Relationship between observations and theoretical model}
\label{sec:relationship}
In order to use observed data to measure the parameters of the whole system, we have to convert the measurements in the theoretical model in the Cartesian coordinates that originated at the black hole center to the real observed data form, RA and DEC offsets that are relative to Sgr A* in the Equatorial coordinate system which is centered at the Earth.

We first generate the orbit of a star with our mixed Python/Fortran code, and get the star's orbital positions, $\vec r_i^{bh} = \{x_i^{bh}, y_i^{bh}, z_i^{bh}\}$, in the black hole frame. 
Then we transform from a Cartesian coordinates centered at Sgr A* to the Equatorial RA and Dec, or in terms of components
\begin{align}
x_i&=x_i^{bh}+d \cos\alpha_{bh}\sin\delta_{bh}\\
y_i&=y_i^{bh}+d\sin\alpha_{bh}\sin\delta_{bh}\\
z_i&=z_i^{bh}+d\cos\delta_{bh},
\end{align} 
where $d$ is the distance from the Earth to the center of black hole, $\alpha_{bh}$ and $\delta_{bh}$ are the RA and DEC of the black hole, x axis points to the First Point of Aries, and z axis points to the same direction as that in the black hole coordinates. The black hole Cartesian coordinates and the Earth Cartesian coordinates are only a translation of their origins described by $\vec d$. Then we convert the positions of the star from Cartesian coordinates centered at the Earth to the Equatorial coordinates,
\begin{align}
\alpha_i&=\arctantwo(y_i,x_i)\\
\delta_i&=\sin^{-1}{\frac{z_i}{\sqrt{x_i^2+y_i^2+z_i^2}}},
\end{align}
where $\alpha_i$ is zero in the x-axis direction, and increases to $2\pi$ along the celestial equator counterclockwise as viewed from the North pole, and $\delta_i$ is zero in the celestial equator, positive to the north and negative to the south of the celestial equator.
We subtract from $\{\alpha_i, \delta_i\}$ a reference position such as the astrometry position $\{\alpha_0=\text{17H43M02S},\ \delta_0=-28.7944^\circ\}$ \citep{cfa-sgr} of Sgr A*, and get the observed RA and DEC offsets $\{\Delta\alpha_i, \Delta\delta_i\}$ relative to Sgr A*, similar to those in the observed data, where $\Delta \alpha_i = \alpha_i - \alpha_0$ and $\Delta \delta_i = \delta_i - \delta_0$. 
Note that the values of $\{\alpha_0, \delta_0\}$ for which we use in this paper have been fine-tuned over the years \citep{sgr-2006}, but those values do not affect our study results because the observables are relative sky locations to $\{\alpha_0, \delta_0\}$, not absolute positions. As long as the measurements are always relative to the same object, it does not even matter whether we take Sgr A* as the reference. Notice that the position of Sgr A* does not necessarily co-locate the center of the black hole. The difference between them can be modeled with five parameters, including the relative position of the black hole to the Sgr A*, $\Delta\alpha_{bh}$ and $\Delta\delta_{bh}$, and the uniform RA and DEC velocities and radial velocity, $\{v_{\alpha_{bh}}, v_{\delta_{bh}}, v_{r,bh}\}$, of the black hole relative to the Sgr A*.
The radial velocities of the  stellar orbit are evaluated as $v_{r,i} = \vec v_i\cdot \hat r_i$, where $\hat r_i=\vec r_i/r_i$ are the unit vectors of line of sight.

Based on our model, the following parameters are measured from the data: the six orbital parameters of the star $\{a, e,
\Phi_0, \beta, \gamma, \psi \}$ (where $a$ is semimajor axis, $e$ is eccentricity, $\Phi_0$ is the initial orbital
phase at some moment, and the other three are Euler angles following a z-x-z definition), the three black hole spin
components ${\bf {J}}=\{J_x, J_y, J_z\}$ in Cartesian coordinates or ${\bf {J}}=\{J, \phi_J, \theta_J\}$ in Spherical coordinates as what we used in the code, the mass of the black hole $M$, the position of the black hole relative to the
Sgr A*  $\{d, \Delta\alpha_{bh}, \Delta\delta_{bh}\}$ (where $d$ is the
distance from Sgr A* to us and the other two indicate the black hole's astronomical position). We ignore the motion of the
black hole relative to the Sgr A* $\{v_{\alpha_{bh}},v_{\delta_{bh}}, v_{r,bh}\}$. To test the no-hair theorem, we also
use two more parameters, the quadrupole term, $Q_2$, of the black hole potential, and the quadrupole term, $Q_X$, due to the external potential of an intermediate-mass black hole or other S-stars outside S2's orbit. Those two parameters can be combined into one parameter, the quadrupole term $Q$, where $Q=Q_2+Q_X$. Throughout the paper everything is in the units of $M_*=4.00\times 10^6\ M_\odot$ when we perform calculations.

\section{Measuring parameters}
\label{sec:Fisher}

\subsection{Bayesian formalism}
\label{subsec:measuring-bayesian}
According to the Bayesian paradigm, a prior distribution $p(\vec\lambda)$ is used to quantify our knowledge about a set of unobservable parameters $\vec\lambda$ in a statistical model when no data are available. We can update our prior knowledge using the conditional distribution of parameters, given observed data $D$, via the Bayes theorem. Suppose that the likelihood, or the distribution of the data from an assumed model that depends on the parameter $\vec\lambda$ is denoted by $p(D|\vec\lambda)$, Bayes theorem updates the prior to the posterior by accounting for the data,

\begin{align}
p(\vec\lambda|D)=\frac{p(D|\vec\lambda) p(\vec\lambda)}{p(D)},
\end{align}
where $p(D)=\int p(D|\vec\lambda) p(\vec\lambda) d\vec\lambda$ is the evidence of the data and also a normalizing constant for the same model.

To separate issues pertaining to measurements from physics from simplified models of stellar orbits, we describe results
using the real observation scenario, where only the angular offsets and radial velocity can be measured. Note that for comparison and to validate our MCMC method, we also employ idealized theoretical measurement scenarios in Appendix \ref{sec:toyincartesian}.  
This realistic measurement model accounts for all of the parameters described in Section \ref{sec:relationship}.  The probability distribution of the data given parameters $\vec\lambda$ is
\begin{align}\label{eqn:likelihood}
p(D|\vec\lambda) &=
 \prod\limits_{k}^{N_{\Delta\alpha}}(2\pi \sigma_{\Delta\alpha_k}^2)^{-1/2} \exp - \frac{[\Delta\alpha(t_k|\vec\lambda)  - \Delta\alpha_k]^2}{2\sigma_{\Delta\alpha_k}^2}
\nonumber \\
 &\times\prod\limits_{k}^{N_{\Delta\delta}}(2\pi \sigma_{\Delta\delta_k}^2)^{-1/2} \exp - \frac{[\Delta\delta(t_k\vec|\lambda)  - \Delta\delta_k]^2}{2\sigma_{\Delta\delta_k}^2}
\nonumber \\
&\times \prod\limits_{k}^{N_{v_r}} (2\pi \sigma_{v_{r,k}}^2)^{-1/2} \exp - \frac{[v_r(t_k|\vec\lambda) - v_{r,k}]^2}{2\sigma_{v_{r,k}}^2},
\end{align}
where $\Delta\alpha(t_k|\vec\lambda)$ and $\Delta\alpha_k$ are the theoretical prediction of the RA offset and the observation, respectively, at epoch $t_k$. The notations are similar for the other two observables, i.e., the DEC offset and the radial velocity. The quantities $\{\sigma_{\Delta\alpha_k}, \sigma_{\Delta\delta_k}, \sigma_{v_{r,k}}\}$ are the measurement uncertainties for the observation at $t_k$. The number of measurements for the three observables are denoted as $N_{\Delta\alpha},\ N_{\Delta\delta},\ \text{and}\ N_{v_r}$, respectively. In the equation above, we have assumed that each measurement of each observable has a noise of Gaussian distribution.


To determine the model parameters and their uncertainties, we use a Markov chain Monte Carlo analysis to sample the likelihood function in Eq.~\eqref{eqn:likelihood}. Specifically, we use an ensemble sampler for MCMC named EMCEE \citep{emcee0, emcee}. \\

\subsection{Fisher matrix}\label{subsec:fishermatrix}

To better understand and validate our MCMC results, and to make efficient projections about future hypothetical measurements, we perform a semi-analytic calculation that approximates the likelihood in Eq.~\eqref{eqn:likelihood} by a locally quadratic approximation.  The coefficient of the second-order term is known as the Fisher matrix.

The illustration of the mechanics of a Fisher matrix calculation is shown in Appendix \ref{sec:toyincartesian} by employing an idealized measurement model in Cartesian coordinates. 
For the observations, we can do the same by exploiting in the special case that the observed data is exactly as predicted by some
set of model parameters $\vec\lambda'$, i.e., $\Delta\alpha_k  = \Delta\alpha(t_k|\vec\lambda')$, $\Delta\delta_k  = \Delta\delta(t_k|\vec\lambda')$, and $v_{r,k} =  v_r(t_k|\vec\lambda')$.  Using a first-order Taylor series
expansion $\Delta\alpha(t_k|\vec\lambda)-\Delta\alpha(t_k|\lambda')\simeq \delta \lambda^a \partial \Delta\alpha(t_k)/\partial \lambda_a$ for
the RA offset $\Delta\alpha$ versus parameters $\vec\lambda$ (here $\lambda^a$ are the elements of $\vec\lambda$ and the same index $a$ means contraction) and similar for the other two observables, we find that the conditional probability of the data given $\vec\lambda$ can be approximated by
\begin{align}
\ln p(D|\vec\lambda)  &= \text{const} - \frac{1}{2} \Gamma_{ab} \delta \lambda_a \delta \lambda_b
\end{align}
with
\begin{widetext}
\begin{align}\label{eqn:fishermatrix}
\Gamma_{ab} &= \sum_k \left[\frac{C_{\lambda_a, \Delta\alpha_k} C_{\lambda_b, \Delta\alpha_k}}{\sigma_{\Delta\alpha_k}^2}+\frac{C_{\lambda_a, \Delta\delta_k} C_{\lambda_b, \Delta\delta_k}}{\sigma_{\Delta\delta_k}^2}+\frac{C_{\lambda_a, v_{r, k}} C_{\lambda_b, v_{r, k}}}{\sigma^2_{v_{r, k}}}\right],
\end{align}
\end{widetext}
where $\Gamma_{ab}$ is the Fisher matrix. For a parameter in $\vec\lambda$ that has two values $\lambda_a$ and $\lambda_a^\prime$ with $\delta\lambda_a$ difference that results in two orbits, the components in Eq.~\eqref{eqn:fishermatrix} for this parameter are

\begin{eqnarray}
C_{\lambda_a, \Delta\alpha_k}\equiv\frac{\partial\Delta\alpha(t_k)}{\partial \lambda_a} = \frac{\Delta\alpha(t_k|\lambda_a) -\Delta\alpha(t_k|\lambda_a^\prime)}{\delta \lambda_a}
\label{componentRA}
\end{eqnarray}

\begin{eqnarray}
C_{\lambda_a, \Delta\delta_k} \equiv\frac{\partial\Delta\delta(t_k)}{\partial \lambda_a} =\frac{\Delta\delta(t_k|\lambda_a) -\Delta\delta(t_k|\lambda_a^\prime)}{\delta \lambda_a}
\label{componentDEC}
\end{eqnarray}

\begin{eqnarray}
C_{\lambda_a, v_{r, k}}\equiv\frac{\partial v_r(t_k)}{\partial \lambda_a} = \frac{v_r(t_k|\lambda_a) - v_r(t_k|\lambda_a^\prime)}{\delta \lambda_a}.
\label{componentRadV}
\end{eqnarray}

Having estimated the Fisher matrix and hence approximated $p(D|\vec\lambda)$ by a Gaussian, we can further construct
marginalized distributions for subset variables $\lambda_A$ in $\vec\lambda=(\lambda_A,\lambda_a)$ by integrating out the variables
$\lambda_a$.   In the Gaussian limit, this integration implies the marginalized distribution has a co-variance matrix
$\bar{\Gamma}_{AB}$ given by 
\begin{eqnarray}
\label{eq:Fisher:Marginalize}
\bar{\Gamma}_{AB} =\Gamma_{AB} - \Gamma_{Aa}[\Gamma^{-1}]_{ab}\Gamma_{bB}.
\end{eqnarray}
Because the second term is negative, the marginalized distribution is always wider: additional uncertain degrees of freedom lead to less accurate constraints.

The Fisher matrix is a cross check for the parameter estimations obtained from MCMC. Drawing in the best-fit parameters, the Fisher matrix can give the estimates of the uncertainties of parameters in a few seconds, whereas it
takes MCMC several hours in our problem. A Fisher matrix can also let us test how sensitively the measurement accuracy and hypothesis tests depend on the stellar parameters.


As an illustration of the usefulness of the Fisher matrix, we show in \ref{sec:measureblackholespin} in a concrete scenario the measurement uncertainty of the spin with both a synthetic stellar orbit similar to S2 orbit and one similar to that of S102/S55.  \\

\subsection{Results on the observed data}
\label{subsec:keck-vlt-results}

After testing the validity of our mixed Python/Fortran code using a highly idealized measurement scenario (see Appendix \ref{ap:toy}), we use observed data to measure the parameters of S2 orbit and the properties of the Galactic Center black hole as also reported elsewhere \citep{2008ApJ...689.1044G,2009ApJ...692.1075G}. Our results agree with their work within systematic and statistical errors. This shows that our code works well with observations and therefore the validity of using it is assured to calculate several hypotheses in Section \ref{sec:hypothesestests}.

Keck S2 data \citep{2008ApJ...689.1044G} are used to estimate the parameters assuming the black hole is not free to move relative to us. The modes of the parameters and their 1-$\sigma$ uncertainties are shown in Table \ref{tab:Keck-VLT-S2}.  VLT S2 data in \citep{2009ApJ...692.1075G} are also used to estimate the parameters, see Table \ref{tab:Keck-VLT-S2}. Our parameter estimation results are consistent with Ghez’s and Gillessen’s analyses within $2\sigma$ and the uncertainties are consistent too.
We evaluate how good a model fit is with the chi-square $\chi^2_{\text{dof}}$ statistics. The reduced chi-square value $\chi_{\text{dof}}^2$ is the chi-square value divided by the number of degree of freedom, which is the degree of freedom of the data subtracted by the number of parameters of the model. Notice that for two measurements that were taken at the same time, the mean of the two measurements of $\{\Delta\alpha_i, \Delta\delta_i\}$ is used as the measurement that happened at that time and the larger error bars are used as the measurement uncertainties of the observables. 

\begin{table*}[!htp]
\centering
\caption{Orbital parameters for S2 and the black hole properties with Keck data and VLT data}
\label{tab:Keck-VLT-S2} 
\begin{tabular}{l*{6}{l}l}
\hline
\hline
Parameter	 (Symbol) [Unit]\ & Keck  &  VLT & VLT w/o 2002 \\
\hline
Semimajor axis ($a$) [AU]		&  $980\pm 17.6$ & $1054\pm 17.8$ & $981 \pm 23.8$\\
Eccentricity ($e$) 	&  0.9048 $\pm$0.0038 & 0.8953$\pm$0.0040 & 0.9038 $\pm$0.0060\\
Initial phase ($\Phi_0$) [radian]          & 3.178$\pm$ 0.0029  &3.031 $\pm$ 0.0032 & 3.038 $\pm$ 0.0040\\
Euler angle 1 ($\beta$) [radian] 	& 0.268 $\pm$0.008 & 0.186 $\pm$ 0.0076 & 0.227 $\pm$ 0.0136\\
Euler angle 2 ($\gamma$) [radian] 		& 1.464 $\pm$ 0.013 & 1.490 $\pm$ 0.016 & 1.444 $\pm$ 0.0252\\
Euler angle 3 ($\psi$) [radian]	   & 3.936 $\pm$0.013  & 4.047 $\pm$ 0.012 & 4.030 $\pm$0.0120 \\
Distance ($d$) [kpc]  & 7.328 $\pm$ 0.17 & 8.422 $\pm$ 0.288 & 7.571 $\pm$0.382 \\
RA offset of BH ($\Delta\alpha_{bh}$) [radian]	 & 1.4166 $\times 10^{-8}$ $\pm$4.42 $\times 10^{-9}$ & 4.99$\times 10^{-9}$ $\pm$ 3.15
$\times 10^{-9}$ &9.86 $\times 10^{-9}$ $\pm$3.34$\times 10^{-9}$\\
DEC offset of BH ($\Delta\delta_{bh}$) [radian] & -4.2962 $\times 10^{-8}$ $\pm$ 6.543$\times 10^{-9}$ &-1.84$\times 10^{-8}$ $\pm$
7.41$\times 10^{-9}$ &-1.575$\times 10^{-8}$ $\pm$1.026$\times 10^{-8}$\\
Mass ($M$) [$10^6\ M_\odot$]		& 4.468 $\pm$ 0.236 & 4.492 $\pm$0.244 & 3.624 $\pm$ 0.272\\
Spin ($\bf{J}$) & not measurable & not measurable & not measurable  \\
Reduced chi-square $\chi_{\text{dof}}^2$ [1] &1.4 & 1.0 &1.0 \\
\hline
\hline
\end{tabular}\\
\vspace{0.2 in}
\raggedright
The table shows the estimated modes and 1-$\sigma$ errors of the six parameters of S2 orbit and the seven parameters of the Galactic Center black hole from Keck and VLT data using our MCMC code. These estimations are are consistent with Ghez's \citep{2008ApJ...689.1044G} and Gillessen's \citep{2009ApJ...692.1075G} analyses within $2 \sigma$. The second, third, and fourth rows use Keck data, VLT data, and VLT data subtracted by its data in 2002 to compare with Keck data because Keck does not contain observations in 2002, respectively. The spin of the black hole is not testable with the two data sets. 
\end{table*}

\section{Testing various Hypotheses}\label{sec:hypothesestests}

\subsection{Bayesian hypothesis selection}
We assume that the observed orbital data $D$ to have arisen under one of the two hypotheses ${\bf\mathcal H}_0$ and $\mathcal H_1$ according to probability density $p(D|\mathcal H_0)$ or $p(D|\mathcal H_1)$ and for given prior probabilities $p(\mathcal H_0)$ and $p(\mathcal H_1)=1-p(\mathcal H_0)$, we obtain from Bayes's theorem
\begin{eqnarray}
p(\mathcal H_i|D)=\frac{p(D|\mathcal H_i)p(\mathcal H_i)}{p(D|\mathcal H_0)p(\mathcal H_0)+p(D|\mathcal H_1) p(\mathcal H_1)},  \\
\nonumber & (i=0,1)
\end{eqnarray}
and
\begin{eqnarray}
\frac{p(\mathcal H_0|D)}{p(\mathcal H_1|D)}=\frac{p(D|\mathcal H_0)}{p(D|\mathcal H_1)}\frac{p(\mathcal H_0)}{p(\mathcal H_1)},
\end{eqnarray}
where we define the Bayes factor as
\begin{eqnarray}
B_{01}=\frac{p(D|\mathcal H_0)}{p(D|\mathcal H_1)}.
\label{k_hypothesis}
\end{eqnarray}
When the two hypotheses are equally probable, the Bayes factor $B_{01}$ is equal to the posterior odds in favor of $H_0$.

If for $\mathcal H_0$ and $\mathcal H_1$ we choose models $\mathcal M_0$ and $\mathcal M_1$ parametrized by model parameter vectors ${\bf\theta}_0$ and ${\bf\theta}_1$, we then have to select between the two models using the Bayes factor,

\begin{eqnarray}
B_{01}=\frac{p(D|\mathcal M_0)}{p(D|\mathcal M_1)}=\frac{\int p({{\bf\theta}_0}|\mathcal M_0)p(D|{{\bf\theta}_0},\mathcal M_0) d{{\bf\theta}_0}}{\int p({{\bf\theta}_1}|\mathcal M_1)p(D|{{\bf\theta}_1},\mathcal M_1) d{{\bf\theta}_1}},
\label{eq:bayesfactor}
\end{eqnarray}
where $p({{\bf\theta}_i}|\mathcal M_i)$ is the prior probability distribution function of parameter vector ${\bf\theta}_i$ in $\mathcal M_i$ for $i=0,1$.

\subsection{Measuring the black hole spin}
\label{sec:measureblackholespin}

\subsubsection{Does the Galactic Center black hole spin?}
\label{sec:blackholespintest}
Measuring the spin of the Galactic Center black hole is of significant interest. Short of an accurate measurement, one can assess the evidence of the existence of any spin. 
Working in the framework of general relativity, we choose the same parameter vector, except the spin, for both the non-spin model ($\mathcal M_0$) and the spin model ($\mathcal M_1$) that address the S2 orbit around the Galactic Center black hole, i.e., ${{\bf\theta}_0}=\{a, e, \Phi_0, \beta, \gamma, \psi, d, \Delta\alpha_{bh}, \Delta\delta_{bh}, M\}$ and ${{\bf\theta}_1}=\{a, e, \Phi_0, \beta, \gamma, \psi, d, \Delta\alpha_{bh}, \Delta\delta_{bh}, M,  {\bf J}\}$, and apply Bayesian statistics to answer the question.

Our models $\mathcal M_0$ and $\mathcal M_1$ are nested, i.e., $\mathcal M_1$ reduces to $\mathcal M_0$ when the spin $J$ or dimensionless spin $\chi$
acquires 0. For a smooth, marginalized posterior probability distribution  $P(J,\mathcal M_1|D)$ of spin $J$ for model $\mathcal M_1$ that is
obtained from an MCMC sampling and has a maximum, we define the $68.3\%$ credible interval to be $\chi\in[\chi_L, \chi_H]$ such that
$\int_{\chi_L}^{\chi_H} P(\chi,\mathcal M_1|D) d \chi=0.683$ with $P(\chi_L,\mathcal M_1|D)=P(\chi_H,\mathcal M_1|D)$. 
For Keck data of S2 orbit up to 2007 in Table 3 in \citep{2008ApJ...689.1044G} and VLT data up to 2009 in
\citep{2009ApJ...692.1075G}, respectively, we use our MCMC code to obtain the posteriors for dimensionless spin $\chi$ under the spin model $\mathcal M_1$, and both of the spin posteriors are uniform. 
It is uninformative about the spin of the Galactic Center black hole with either data set.

The Bayes factor $B_{01}$ is evaluated for the selection of our two models with Keck's S2 data. Parameter estimation is done on the non-spin model $\mathcal M_0$ too. The Bayes factor $B_{01}$ in favor of $\mathcal M_0$ than $\mathcal M_1$ is 1.2, which is calculated from Eq.~\eqref{eq:bayesfactor} with the posteriors from the MCMC samplings with the Keck data for the two models that represent the two hypotheses. As stated at the beginning of this section, the parameters for the two models are the same except that there is no spin parameter $\bf{\text{J}}$ in $\mathcal M_0$. The value $B_{01}=1.2$ is interpreted as that the non-spin model $\mathcal M_0$ is slightly (but barely worth mentioning) more strongly supported by the data than the spin model $\mathcal M_1$. We find that this does not contradict with the most recent constraints on the Galactic Center black hole spin \citep{Fragione:2020khu} where their estimate is $\chi \leq 0.1$ strictly.

\subsubsection{Stellar orbit measurement scenarios for decisive constraints on the Galactic Center black hole spin}
\label{sec:blackholespinmeasurementscenarios}

\begin{figure}[ht!]
\includegraphics[width=\columnwidth]{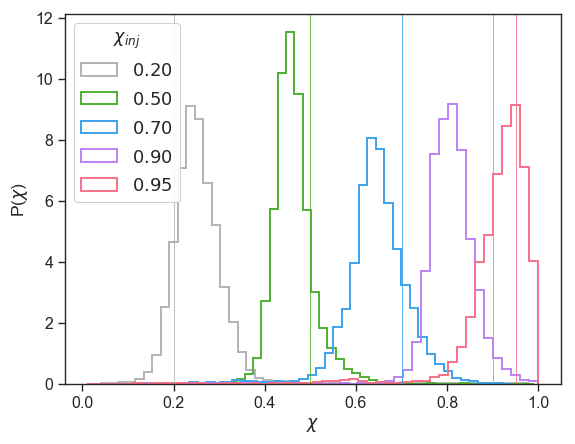}
\includegraphics[width=\columnwidth]{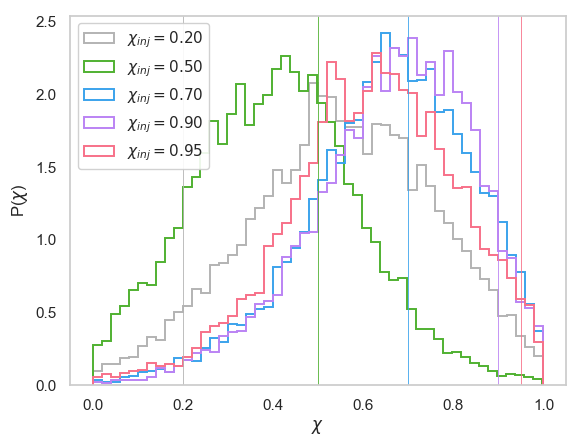}
\caption{\label{fig:constrainspinwithfakedata}The marginalized posteriors of dimensionless spin $\chi=J/M^2$ for the fake observations of S2 in Scenario I (top panel) and II (bottom panel) in Table \ref{tab:S2-like}. The thin vertical lines are injected values. The range of the spin $J$ is $[0,M^2]$, where $M=1.15\ M_*$ and $M_*=4.00\times 10^6\ M_\odot$. The measurement uncertainties $\{\sigma_{\Delta\alpha}=\sigma_{\Delta\delta}=10\ \mu$arcsecond, $\sigma_{v_r}=500$ m/s$\}$ are the limits of GRAVITY at a distance of 8 kpc. The star has an S2-ish orbit. It is observed once per week for 2080 weeks or about 40 years for 2.5 full orbits for the top panel; and twice per week for 1000 weeks or about 20 years for more than one full orbit for the bottom panel.} 
\end{figure}

\begin{figure}[ht!]
\label{90confidence}
\includegraphics[width=\columnwidth]{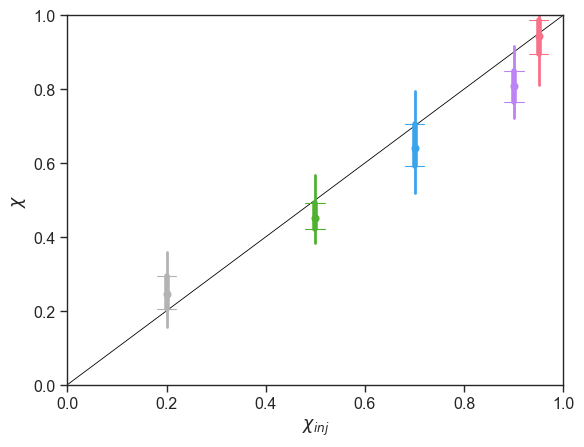}
\caption{\label{fig:spincredibleinterval}The mode and two credible intervals of the dimensionless spin $\chi$ of the Galactic Center black hole as a function of injected dimensionless spin $\chi_{\text{inj}}$ for fake observation Scenario I in Table \ref{tab:S2-like} and from the top panel of Figure~\ref{fig:constrainspinwithfakedata}. Dots show the maxima of posterior estimates of $\chi$; bars indicate the $68.3\%$ ($1\sigma$, thick with caps to the ends) and the $95.4\%$ ($2\sigma$, thin) credible intervals. The black thin line is when $\chi=\chi_{\text{inj}}$. 
}
\end{figure}

What measurements of star S2 can enable us to constrain the black hole spin? We assume a set of future achievable measurement precision $\{\sigma_{\Delta\alpha}=\sigma_{\Delta\delta}=10\ \mu$arcsecond, $\sigma_{v_r}=500$ m/s$\}$, which are the resolution limits of GRAVITY instrument \citep{Gillessen:2010ei,GRAVITY2017}, and use our code to conduct fake/virtual observations of S2 around the Galactic Center black hole and estimate the model parameters including the spin. Specifically, for each simulation we inject a dimensionless spin value $\chi_{\text{inj}}=J_{\text{inj}}/M^2$ to the black hole and let the star  evolve its orbit around this spinning black hole under the equation of motion model in Eq.~\eqref{eq:EOM:SingleBody}. To mimic measurements with noises in them, we add a Gaussian noise of the chosen measurement uncertainty (Table \ref{tab:S2-like}) to the evolved stellar orbit for each observable at  each measurement epoch. Recall that the three observables consist of two angular offsets and one radial velocity. We then use our code to calculate marginalized posterior distributions of the parameters, including the black hole spin, given the fake observed orbital data $D_f$. For $\chi$ it is $P(\chi|D_f)\propto P(\chi)P(D_f|\chi)$. The prior in $\chi$ is uniform for $\chi \in [0,1]$. 

Our virtual observations that can measure the black hole spin with S2 are called Scenario I and summarized in Table \ref{tab:S2-like}. The stellar orbit has $a=2.65\times 10^{4}\ M_*=1060\ \text{AU}\approx 5\ \text{mpc},\ e=0.8847,\ \Phi_0=-0.1$ (which corresponds to Aug, 2017 for S2, and we assume that is when we start the virtual observations) and three Euler angles that have the values of an S2 orbit. The parameters for the black hole are $M=4.6\times10^6\ M_\odot=1.15\ M_*$ and its sky position $d=8.0$ kpc, $RA=265.75^\circ$ and $DEC=-28.79^\circ$ which are determined by the observed S2 orbit in Table \ref{tab:Keck-VLT-S2}. The injected dimensionless spin $\chi_{\text{inj}}$ values are ${\{0.2,\ 0.5,\ 0.7,\ 0.9,\ 0.95\}}$ and the spin direction is randomly chosen as $\{\phi_J=\pi,\ \cos\theta_J=0.2\}$ for any of those five injected spin magnitudes.  We get from MCMC the marginalized posterior $P(\chi|D_{f})$ for the fake observed data with that injected spin value $\chi_{\text{inj}}$. We do the same for different injected black hole spin values. The fake observations are conducted once per week for 2080 weeks, or 40 years, about two and a half complete orbits in Scenario I. In Scenario II, we take the same number of measurements but the measurements are arranged twice frequently during half of the observing time compared to Scenario I. 

\begin{figure}[ht!]
\includegraphics[width=\columnwidth]{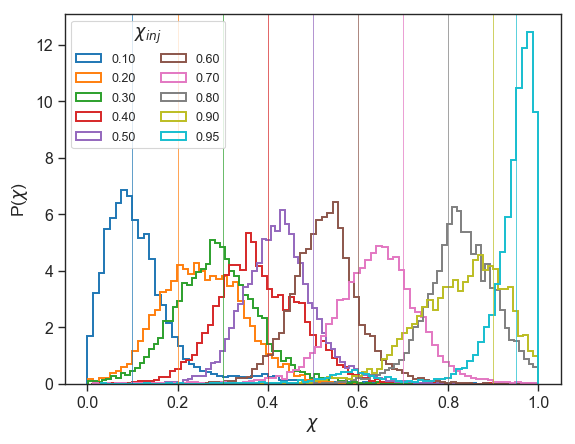}
\caption{\label{fig:constrainspinwithfakedatahalfS2}The marginalized posteriors of the dimensionless spin $\chi$ for different injected values $\chi_{\text{inj}}\in[0,1]$ in Scenario III (see Table \ref{tab:scenarios-summary}). The vertical thin lines are injected values.  The star has an S2-ish orbit except that its orbit is half-sized. It is observed once per week for 800 weeks.}  
\end{figure}

The plots in Figure \ref{fig:constrainspinwithfakedata} show the marginalized posteriors of $\chi$ for different injected values $\chi_{\text{inj}}$ for Scenario I (top panel) and Scenario II (bottom panel). Even though they have the same number of data points and the observation are done on the same star S2, Scenario I has better constraints on the spin than the Scenario II. The minimum we should do to be able to constrain the black hole spin with S2 orbit during 40 years, or a person's entire academic career, is to observe it once per week and record the three orbital observables. Observing less frequently or for shorter amount of time will not enable us to constrain the black hole spin decisively. 

The parameter estimation results in Scenario I are also illustrated on a recovered v.s. injected plot. Figure~\ref{fig:spincredibleinterval} shows the credible intervals of the black hole spin as a function of injected spin value. For each injected value, we plot two error bars. The thick-lined error bar with caps to the two ends is the $68.3\%$ credible interval and the thin-lined error bar without caps is the $95.4\%$ credible interval of the marginalized posterior that corresponds to its injected dimensionless spin $\chi_{\text{inj}}$. The big round dot of the same color in that bar is the value of maximum posterior. The uncertainty of the measured spin is about $0.1$ at the $68.3\%$ credible interval for 40 years of weekly measurements of S2 orbit with GRAVITY's best resolution at the Galactic Center.

On the other hand, if we can find a star that is closer to the Galactic center, then it is possible to sooner achieve a similar measurement precision on the spin as in Scenario I. The possible existence of such stars has been studied in previous research on the stellar density distribution around an isolated massive black hole. Stars can get almost as near as a few hundreds of the Schwarzschild radii of the supermassive black hole through the mechanisms of binary disruptions and dynamical relaxation \citep{Hopman:2006aj,2017ARA&A..55...17A,2018ApJ...852...51F}. The recently found faint stars of shorter periods than S2 \cite{2020ApJ...899...50P} also indicate that there could be stars even closer to the Galactic Center. In Scenario III, we consider a star that orbits around the black hole on half the size of the S2 orbit. We also observe it weekly for $\sim2.5$ full orbits or 800 weeks, see Table \ref{tab:scenarios-summary}. Figure \ref{fig:constrainspinwithfakedatahalfS2} shows the posteriors of $\chi$ for Scenario III. Note that for all the cases in the Scenarios I through VI, the reduced least-square value is $\chi^2_{\text{dof}}\approx 1$ which means the sampling is converged.

\begin{table*}[ht]
\centering
\caption{Scenario I: Fake Observation of S2}
\label{tab:S2-like} 
\begin{tabular}{l*{5}{l}l}
\hline
\hline
Parameter (Symbol) [Unit]\ & Injected parameter value \\
\hline
Star & S2 \\
Semimajor axis ($a$) [AU]		&  $1060$ \\
Eccentricity ($e$) 	&  0.8847 \\
Initial phase ($\Phi_0$) [radian] & -0.100 (Aug 2017)\\
Euler angle 1 ($\beta$) [radian] 	& 0.169\\
Euler angle 2 ($\gamma$) [radian] 		& 1.515\\
Euler angle 3 ($\psi$) [radian]	   & 4.046 \\
Dimensionless spin ($\chi_{\text{inj}}$) & $\{0.2, 0.5, 0.7, 0.9, 0.95\}$ & &  &\\
Spin angle 1 ($\phi_J$) & $\pi$ \\
Spin angle 2 ($\cos\theta_J$) &0.200  \\
Quadrupole moment $Q_2$ & $-J_{\text{inj}}^2/M$ with $J_{\text{inj}}=\chi_{\text{inj}}M^2$\\ 
Mass ($M$) [$M_\odot$] & $4.60\times 10^6$ \\
Distance ($d$) [kpc]  & 8.00 \\
RA of BH ($\alpha_{bh}$) [degree]	 & 265.754795 \\
DEC of BH ($\delta_{bh}$) [degree] & -28.794375 \\
Measurement uncertainty in RA offset ($\sigma_{\Delta\alpha}$) [$\mu$arcsecond] & $10$\\
Measurement uncertainty in DEC offset ($\sigma_{\Delta\delta}$) [$\mu$arcsecond] & $10$\\
Measurement uncertainty in radial velocity ($\sigma_{v_r}$) [m/s] & 500 \\
Orbital period (T) [week] & 823 \\
Measurements & 2080 weekly\\
\hline
\hline
\end{tabular}\\
\vspace{0.2 in}
\raggedright
The table lists the injected parameters and observation strategy for the fake observations of star S2 in Scenario I.  The posteriors from parameter estimation are shown in the top panel of Figure \ref{fig:constrainspinwithfakedata}. Here $M_*=4\times 10^6\ M_\odot$ is used as a scale of the black hole mass.
\end{table*}

\begin{table*}[ht]
\centering
\caption{Summary of Fake Observation Scenarios on S2 and future stars}
\label{tab:scenarios-summary} 
\begin{tabular}{l*{5}{l}l}
\hline
\hline
Scenario  & Star & Semimajor axis [AU] & Injected spin $\chi_{\text{inj}}$ & Period [week] & Measurements \\
\hline
Scenario I & S2 &  $1060$ & $\{0.2, 0.5, 0.7, 0.9, 0.95\}$ & 823 & 2080, weekly\\
Scenario II	& S2 &  $1060$ & $\{0.2, 0.5, 0.7, 0.9, 0.95\}$ & 823 & $1040\times 2$, semiweekly\\
Scenario III & Star of half S2 orbit & $530$ & $\{0.1, 0.2, ..., 0.8, 0.9, 0.95\}$ & 291&  800, weekly \\
Scenario IV & S2 & $1060$  & $\{0.7\}$ & 823 & $2080 \times 7$, daily\\ 
Scenario V & Star of one fifth S2 orbit & $212$ & $\{0.2, 0.5, 0.7, 0.9, 0.95\}$ & 73.6 & $1040$, weekly\\
Scenario VI & S102/S55-ish & $920$ & $\{0.9\}$ & 665 & 2080, weekly\\
\hline
\hline
\end{tabular}\\
\vspace{0.2 in}
\raggedright
Summary of the differences among the four fake observation scenarios. For the parameters that are not specified for Scenarios II, III, IV, V, and VI here, they take the same values as in Scenario I in Table \ref{tab:S2-like} except that for Scenario VI the eccentricity is $e=0.721$.  The estimated posteriors are shown in the top and the bottom panels of Figure \ref{fig:constrainspinwithfakedata} for Scenario I and II, Figure \ref{fig:constrainspinwithfakedatahalfS2} for Scenario III, Figure \ref{fig:onefifthS2orbitQ2test} for Scenario V, and Figure  \ref{fig:compareS2andS102} for Scenario VI. 
\end{table*}

\subsubsection{Analytic expression for the measurement uncertainty of Galactic Center black hole spin}
\label{sec:blackholespinmeasurementFishermatrix}
We also develop a convenient analytic expression from Fisher matrix for the measurement uncertainty of the black hole spin with generic observation strategies on stellar orbits. From Eq.~\eqref{eq:FisherMatrixSpinComponentforGenericOrbits}, we can derive that the uncertainty in the spin measurement $\sigma_J$ (or $\sigma_{\chi}$ for the dimensionless spin), i.e., the inverse square root of $\Gamma_{J_aJ_b}$, is determined by
\begin{eqnarray}
\label{eq:SpinUncertaintyScaling}
\sigma_{\chi}\propto\sigma_{J} \propto\frac{ a^2\sigma_r}{\sqrt{N}T }\frac{ (1-e^2)^{\frac{3}{2}}} {(13e^4+9e^2+3)^{\frac{1}{4}}},
\end{eqnarray}
where $a$ is the semimajor axis, $e$ is the eccentricity, $\sigma_r$ is the stellar orbit measurement uncertainty, $T$ is the duration of observation time, and $N$ is the number of measurements. 

\begin{figure}[t!]
\includegraphics[width=\columnwidth]{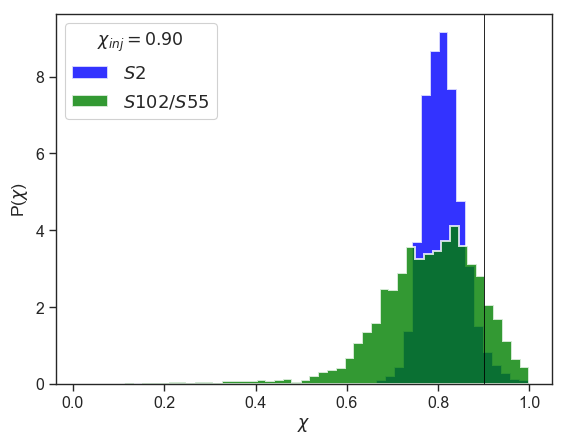}
\caption{\label{fig:compareS2andS102}The marginalized posteriors of the dimensionless spin $\chi$ for injected value $\chi_{\text{inj}}=0.90$ for S2 (blue) in Scenario I in Table \ref{tab:S2-like} and S102/S55-ish (green) in Scenario VI in Table \ref{tab:scenarios-summary}. The vertical thin line is injected value.  Both stars are observed once per week for 2080 weeks or 40 years. The differences of the two orbits are their eccentricities and the semimajor axises, which are taken the values for S2 and S102/S55. }  
\end{figure}

\renewcommand{\arraystretch}{1.}
\begin{table}[ht]
\centering
\caption{How Many Years of Weekly Observations Are Required to Reach $\sigma_{\chi}\sim 0.1$ with S2, S62, S4711 and S4714}
\label{tab:yearsofobservations-mostpromisingstars} 
\begin{tabular}{l*{4}{l}l}
\hline
\hline
Star & $\sigma_r=6.5$ mas & $\sigma_r=0.65$ mas & $\sigma_r=65\ \mu$as & $\sigma_r=10\ \mu$as \\
\hline
S2 &  \NA & \NA & \NA & 40\\
S62 &  410 & 88 & 19 & 5.5 \\
S4711 & 3100 & 670 & 150 & 42\\
S4714 & 300 & 66 & 14 & 4.1\\
\hline
\hline
\end{tabular}\\
\vspace{0.2 in}
\raggedright
The number of years of weekly observations required to reach a black hole spin measurement precision similar to Scenario I (see the top panel of Figure~\ref{fig:constrainspinwithfakedata}). 
The uncertainties in the estimation of numbers of years are less than $10\%$, which are straightforward with Eq.~\eqref{eq:SpinUncertaintyScaling} and the uncertainties of the orbital parameters in Table 1 of \citep{2020ApJ...899...50P}.
Note that mas is milliarcseond and $\mu$as is microarcsecond.
\end{table}

We now use Fisher matrix to check against the MCMC method, with stellar orbits similar to those of S2 and S102/S55 in fake observation Scenarios I and VI, see Figure \ref{fig:compareS2andS102}. In this scenario, both stars are observed with
the same number of measurements $N$, the same stellar orbit measurement accuracy $\sigma_r$, and the same duration of observation $T$. The dimensionless spin uncertainty $\sigma_{\chi}$, is then scaling only to the semimajor axis $a$ and the eccentricity $e$. The closer the orbit and the larger the eccentricity, the more accurate we can constrain the black hole spin. Plugging into the values of those two quantities for the two stars respectively from Tables \ref{tab:S2-like} and \ref{tab:scenarios-summary}, we can obtain the dimensionless spin uncertainty ratio constrained from the orbits of S2-ish to S102/S55-ish using Eq.~\eqref{eq:SpinUncertaintyScaling} for any black hole spin value. It is 0.34 for the two scenarios considered. This is consistent with the MCMC results shown in Figure \ref{fig:compareS2andS102}, where the uncertainty ratio is $0.129/0.267\approx0.48$ 
for $\chi_{\text{inj}}=0.9$.

The above example uses both the Fisher matrix and the MCMC method. 
In general, with Fisher matrix, we know from Eq.~\eqref{eq:SpinUncertaintyScaling} that an orbit with smaller semimajor axis and larger eccentricity can provide better constraints on the black hole spin. From Eq.~\eqref{eq:SpinUncertaintyScaling}, we also know that with the same semimajor axis $a$, which translates to the same orbital period $P$ with Kepler's third law, larger eccentricity (highly eccentric orbits) can constrain the black hole spin more precisely given the same observation strategy. This is because with the same semimajor axis, the pericenter of the more eccentric orbit is closer to the black hole and thus more impacted by the black hole's gravitational potential. This is why S2 can provide better
constraints (as is the case in our simulation shown in Figure~\ref{fig:compareS2andS102}) on the black hole spin than what S102/S55 can do, even though S102/S55 has a shorter period. In addition, by observing $n$ times as often, we can improve the measurement error bars of spin by a factor of $\sqrt{n}$ to guide our numerical simulation in terms of observation strategy choices.

Similarly, from Eq.~\eqref{eq:SpinUncertaintyScaling} we can see that if we cannot observe S2 for 40 years weekly in order to determine the black hole spin, which is most likely, the solution is to find a closer star or improve the instrument measurement precision. If we have $k$ stars whose orbits are similar to S2 and we observe each of them equally frequently and for the same amount of time duration, we are expected to see an improvement of a factor of $\sqrt{k}$ in the measurement uncertainty of black hole spin. 

Another application of Eq.~\eqref{eq:SpinUncertaintyScaling} is on the number of years of weekly observations required to reach a spin measurement uncertainty of about 0.1 with the recently discovered stars S62, S4711, and S4714, with respect to the telescope resolution $\sigma_r$. The resolution used in their discoveries is $6.5$ mas \citep{2020ApJ...899...50P}, but we also list the scenarios where better resolutions are achieved. The eccentricity and semimajor axis values are taken from Table 1 of \citep{2020ApJ...899...50P}. The results are shown in Table~\ref{tab:yearsofobservations-mostpromisingstars}. Note that theoretically for the same measurement resolution, e.g., $\sigma_r=10\ \mu\text{as}$, stars S62 and S4714 can constrain the black hole spin sooner than S2. However, S62 and S4714 are 4 magnitudes fainter than S2 \citep{2020ApJ...899...50P}. This can cause larger uncertainties on the orbital measurements of the former than the latter, and hence lead to a cross-row comparison in Table~\ref{tab:yearsofobservations-mostpromisingstars}.

\subsection{Testing no-hair theorem}\label{sec:nohairtheorem}
According to the black hole no-hair theorem, a black hole is completely characterized by its mass $M$, angular momentum
(or spin) $J$, and charge $q$. For an astrophysical black hole which is electrically neutral, it is fully described by
two quantities, $M$ and $J$. As a consequence, the quadrupole moment $Q_2$ of its external spacetime is given by
$Q_2=-J^2/M$. The quadrupole moment can cause the stellar orbits around the black hole to precess, and the precession rate is on the order of $1\ \mu$arcsecond for a highly eccentric orbit around the Galactic Center supermassive black hole with orbital period of years. This makes it possible to use the stellar orbit data from the modern
infrared telescopes to test the no-hair theorem. In reality, there is perturbing external
quadrupole moment $Q_X$ (see Section \ref{sec:Methods}) due to the S star cluster, dark matter, and intermediate-mass black
holes that are close to the Galactic Center. This should also be taken into consideration. In this study, we employ the most 
optimistic possible scenario, equivalent to perfect knowledge of any external tidal potential.

We first apply our MCMC code to the VLT orbital data of S2 to obtain the marginalized posterior probabilities of spin $J$ and quadrupole moment $Q_2$, and they are both flat. The existing data are not sufficient for us to draw a conclusion on the no-hair theorem. This is not surprising because we cannot even constrain spin yet.

\begin{figure}[t]
\includegraphics[trim=.cm .cm .cm .cm, clip=true, width=0.99\columnwidth]{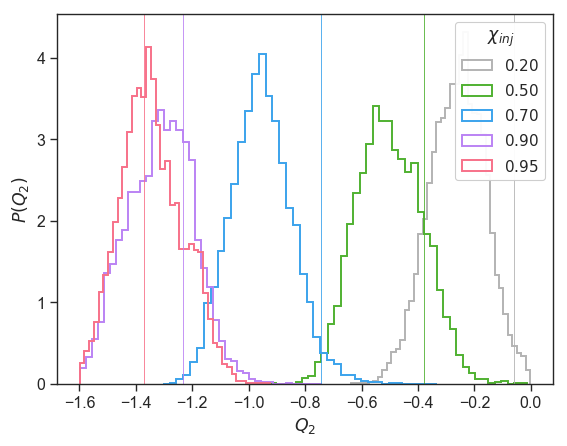}
\includegraphics[trim=.cm .cm .cm .cm, clip=true, width=0.99\columnwidth]{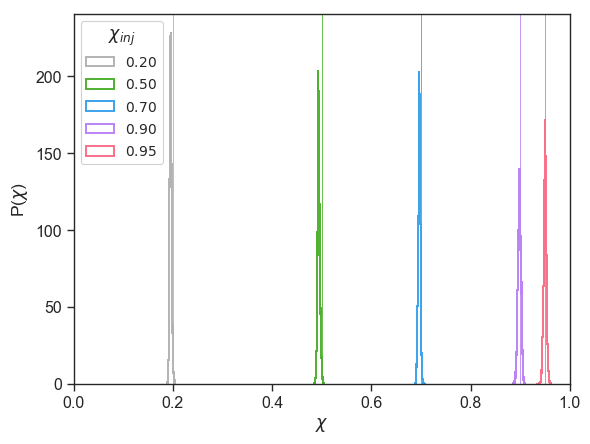}
\caption{\label{fig:onefifthS2orbitQ2test}Posteriors of $Q_2$ and $\chi$ for Scenario V in Table \ref{tab:scenarios-summary}. The thin vertical lines are the injected values. The top panel shows the marginalized posteriors of quadrupole $Q_2$ for different injected values $\chi_{\text{inj}}$ and their corresponding injected quadruple values $Q_{2,\text{inj}}=-\chi_{\text{inj}}^2 M^3$. The bottom panel is for the posteriors of the dimensionless spin $\chi$. The star has an orbit that is one fifth the semimajor axis of S2. It is observed once per week for about seven full orbits. }
\end{figure}

\subsubsection{Can the S2 orbit test no-hair theorem?}
\label{subsubsec:nohairwithS2}

Can any strategies of observations on S2 enable us to test the no-hair theorem in the future? Applying the same method that is used in Section \ref{sec:blackholespinmeasurementscenarios}, we conduct fake observations of S2 again. Specifically, we first do Scenario IV. The injected parameters and the observing strategy can be found in Table \ref{tab:scenarios-summary} and its reference Table \ref{tab:S2-like}. We use our code to generate fake observed orbital data points with noise in them for S2 star around our Galactic Center for an injected black hole spin $\chi_{\text{inj}}=0.7$ and quadruple moment $Q_{2,\text{inj}}=0.648$. We virtually observe the star daily for 40 years. We then use MCMC to obtain the marginalized posterior probability distribution of $Q_2$ for the choice of $\chi_{\text{inj}}$ and $Q_{2,\text{inj}}$ values. In the parameter estimation, $Q_2$ is treated as an independent parameter on $J$ and $M$, so it does not follow $Q_2=-J^2/M$. For this set up, it is equivalent to see how well the quadrupole term $Q$, including the external quadrupole moment $Q_X$, can be constrained. The posteriors $p(Q_2)$ for different injected values are all flat. This means, we cannot constrain the quadrupole term or the no-hair theorem even if we observe S2 daily for 40 years with GRAVITY's resolution limits. 

The 40-years daily measurements of S2 orbit with GRAVITY are not sufficient to constrain the no-hair theorem mainly because even at a periapsis distance of about 120 AU or $2800$ times the mass of the black hole, the star is not close enough to the supermassive black hole to be significantly affected by the black hole's quadrupole moment for the telescopes to observe. In order to use S2 to constrain $Q_2$, we will have to improve our angular measurement precision by at least two orders of magnitude and the radial velocity uncertainty by one order of magnitude from our virtual experiments, compared to GRAVITY's limits.

\subsubsection{Highly eccentric stars at 1 mpc for no-hair theorem}
\label{subsubsec:nohairwithfuturestars}

\begin{figure*}
\includegraphics[width=0.99\textwidth]{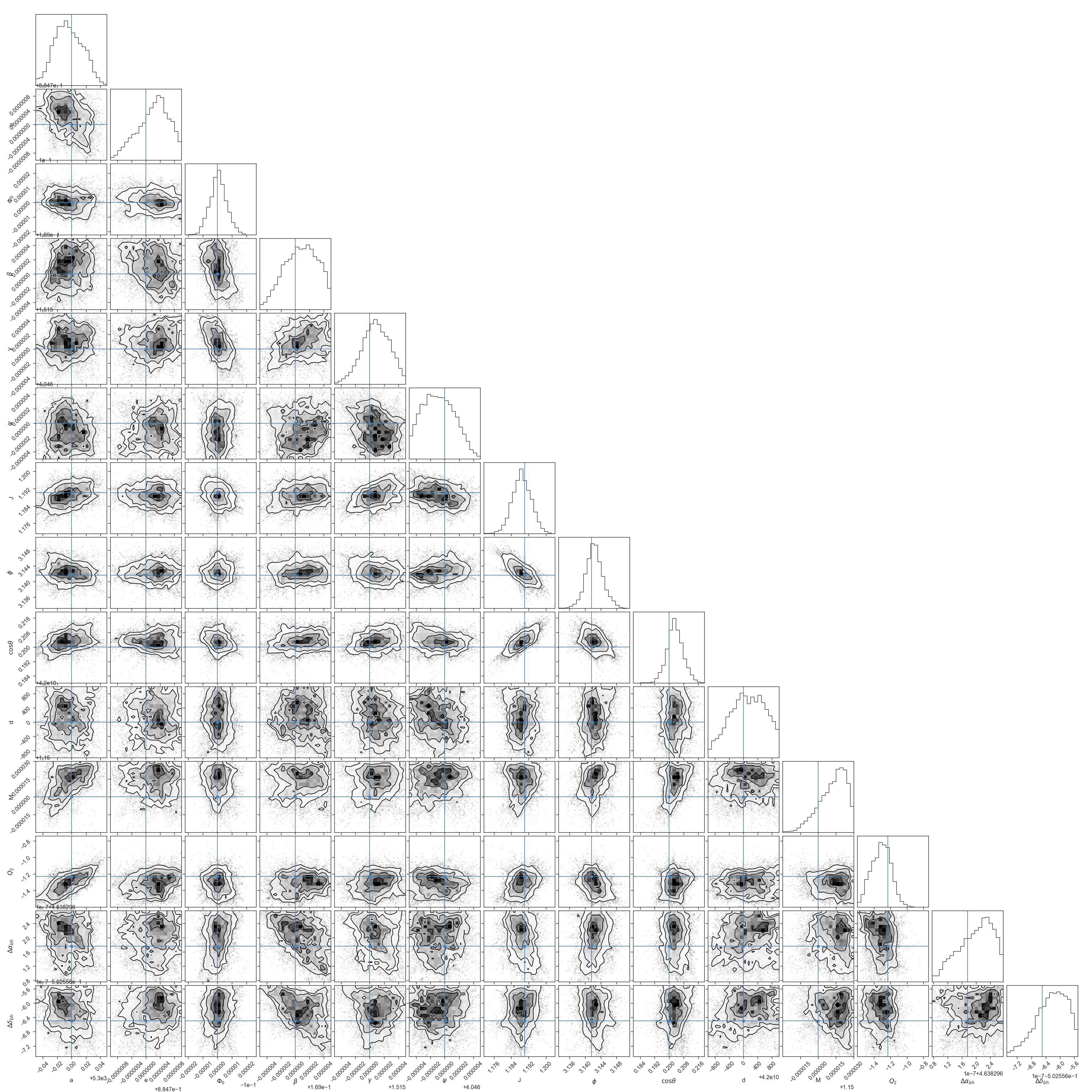}
\caption{Corner plot for the posterior distributions of the parameters for the case of injected $\chi_{\text{inj}}=0.900$ and $Q_{2,\text{inj}}=1.232$ in Scenario V in Table \ref{tab:scenarios-summary}. 
}
\label{fig:corner-q90}
\end{figure*}

Because S2 with even GRAVITY's resolution will not work, we ponder what kind of stellar orbits and observation strategies are needed to test the no-hair theorem then. In Section \ref{sec:blackholespinmeasurementscenarios} we have briefly discussed the possible existence of stars closer to to the Galactic Center than the newly found S62, S4711, and S4714 that are also supported by theories \citep{2017ARA&A..55...17A,2018ApJ...852...51F}. The next generation of extremely large telescopes will discover stars
with orbital periods as small as 1-2 years given their increased sensitivity and angular resolution \citep{Graham:2019mis}. We assume that we are lucky enough to find a star orbiting around the Galactic Center on an orbit that has a fifth of the semimajor axis of S2 ($\sim 200$ AU or 1 mpc). We also assume that all its other orbital parameters including the eccentricity are the same as S2. This star has an orbital period of 73.6 weeks and we observe it once per week for 1040 weeks (20 years and 14 full orbits) in our simulation Scenario V, see Tables \ref{tab:scenarios-summary} and \ref{tab:S2-like}. With this fake observation scenario, we can start to measure the quadrupole moment $Q_2$ for different injected $Q_{2,\text{inj}}$. See the top panel of Figure \ref{fig:onefifthS2orbitQ2test} for the posteriors $P(Q_2)$. In this figure, also plotted is the measurement of dimensionless spin $\chi$ for the various injected values in the bottom panel. We can see that while the $Q_2$ can be measured to a visually distinguishable extent, the black hole spin can be measured at a very high precision, $\sim 0.01$. As an example, we show in Figure \ref{fig:corner-q90} the corner plot of posteriors of all modeled parameters for a specific case, $\chi_{\text{inj}}=0.900$ and $Q_{2,\text{inj}}=-1.232$ in Scenario V. 

\subsubsection{Effects from ambient perturbers}
\label{sec:ambientperturbers}

Now let us discuss how these constraints can be affected by the target star's ambient perturbers, such as a stellar cluster or an intermediate-mass black hole. We first consider the perturbations from other stars within the stellar orbit based on the study in \citep{Merritt:2009ex}, because the small stellar cluster may induce orbital precession of the same order of magnitude as that due to general relativistic effects. The additional stellar cusp has a very small contribution (see their Eq. 10) to the advance of the periapsis but only competes general relativistic effects on long timescales. The vector resonant relaxation could also contribute to the orbital precession. Their Fig.~3 which plots Eq.~30 and Fig.~4 which checks Eqs.~28-30 using an N-body simulation show the transition from the domination by general relativistic effects to that by stellar perturbations under certain conditions. For the distances of interest (semimajor axis about $1\ \text{mpc}$ for the star of one fifth the size of S2 orbit), the stellar perturbation effects should be included only when the stellar cusp consists of mostly $10 M\odot$ black holes or when it consists of solar mass stars but the total mass is over $200 M\odot$ as seen from Fig.~3 of  \citep{Merritt:2009ex}.

We also discuss the effect from a possible massive dark perturber, such as an intermediate-mass black hole. A number of IMBH candidates have been suggested near the Galactic Center \citep{2017IJMPD..2630021M, 2019ApJ...871L...1T, 2020ApJ...890..167T}. An IMBH orbiting around the black hole can cause Kozai oscillations of the orbital parameters of a star if the IMBH is located outside of the stellar orbit. Research with stars and IMBHs at greater distances (50 mpc or larger for a star and 300 mpc or larger for the IMBH) has been nicely performed \citep{2020ApJ...905..169Z}, but it cannot be directly applied to our cases where both the star and the IMBH are much closer to the black hole at 1 mpc level. We use the work in \citep{2009ApJ...705..361G} and focus on the Kozai mechanism. The inner orbit is a star of one fifth the size of S2 orbit with semimajor axis about $a_{\text{in}}=1\ \text{mpc}$. For the Kozai mechanism to operate, the IMBH must lie outside the orbit of the star as illustrated in Section 4.4 of \citep{2009ApJ...705..361G}. For the Kozai effect to dominate, it also requires the the Kozai oscillation period to be smaller than the other effects on the inner orbit. Therefore, the semimajor axis of the IMBH's orbit $a_{\text{out}}$ should be large enough but as small as possible too to have larger and dominating perturbations on the inner orbit. We first consider $a_{\text{out}}=2\ \text{mpc}$ given that the star is highly eccentric ($e\approx 0.9$) and the apoapsis is nearly $2 a_{\text{in}}$. With Eq.~16 and Fig.~10 of \citep{2009ApJ...705..361G}, we can scale the Kozai oscillation period as $T_{K}\approx 0.9\ \text{years}\  (\frac{a_{\text{out}}}{a_{\text{in}}})^2  (\frac{a_{\text{in}}}{\text{mpc}})^{1.5}\frac{M_{\text{BH}}}{M_{\text{IMBH}}}(1-e_{\text{out}}^2)$ 
for a $50^{\circ}$-$85^{\circ}$ relative inclination between the inner and the outer orbits. For $a_{\text{in}}=1\ \text{mpc}$, $a_{\text{out}}=2\ \text{mpc}$, $M_{\text{BH}}=4.5\times10^6\ M_\odot$, $M_{\text{IMBH}}=1000 M_\odot$ and $e_{\text{out}}=0$, the Kozai period is about $1.6\times 10^4$ years. Note that $M_{\text{IMBH}}>1000 M_{\odot}$ at this distance is ruled out by observational constraints in \citep{2009ApJ...705..361G}. The general relativistic precession timescale is 440 years at $a=1\ \text{mpc}$ and $e^2=0.8$ using Eq.~1 of \citep{2009ApJ...705..361G}. Because the Kozai period is much greater than the general relativistic precession timescale, the Kozai effect from the IMBH will be damped by the general relativistic precession for the star in the inner orbit. For the IMBH at larger distances than 2 mpc, the Kozai oscillation period will be even longer and washed out by the general relativistic precession effect. If an IMBH is inside the star's orbit, then it is a discussion similar to that on the inner stellar cluster's effect on the target star of observation. Following the discussion in Section 3 of \citep{2009ApJ...705..361G}, the principal impact of likely nearby IMBHs is on the distribution of S-star orbital parameters over millions of years.  An IMBH can also produce small stepwise perturbations to individual stellar orbits when it passes closely nearby, but this requires fine-tuning and is very unlikely to occur for the few stars expected to be monitored over $\mathcal{O}$(decade) to constrain the central black hole's properties, given a putative IMBH's orbital period is likely comparably long.

\section{Conclusions}
\label{sec:conclusion}
We have introduced a Markov chain Monte Carlo method to constrain the Galactic Center black hole properties with a model of the Galactic Center stellar orbits. We also use Fisher matrix method to check against MCMC when the scenarios allow. Several main conclusions come out of this work. 

First, we conclude that we are not able to constrain the black hole spin or test the no-hair theorem with the existing data of S2 stellar orbit from Keck and/or VLT measurements taken from 1995 to 2007 (2009 for VLT). 

We then provide strategies for future observations on how to constrain the black hole spin with S2 orbit based on simulated fake observation scenarios, assuming future achievable measurement accuracy. With the best measurement resolution by the GRAVITY telescope when it observes at an angular uncertainty of 10 $\mu$arcsecond and a radial velocity uncertainty of 500 m/s, we can constrain the black hole spin at 0.1 precision if S2 is observed once per week for 40 years as shown in Figure~\ref{fig:constrainspinwithfakedata} for Scenario I in Table \ref{tab:S2-like}. 
If we can find a closer star that is half the size of S2 orbit, then we can measure the spin sooner than using S2 as shown in Figure \ref{fig:constrainspinwithfakedatahalfS2} for Scenario III.

We also derive an analytic expression to scale the uncertainty of the spin measurement using Fisher matrix in terms of the observation strategy, the star's orbital parameters, and the instrument precision, see Eq.~\eqref{eq:SpinUncertaintyScaling}. After demonstrating the correctness of this equation with a concrete example labeled Scenario VI and its simulation results shown in Figure~\ref{fig:compareS2andS102}, we continue to apply it to the spin measurement with the recently found stars S62, S4711, and S4714. The numbers of years of weekly observations required to reach a  $\sigma_{\chi}\sim 0.1$ uncertainty are shown in Table~\ref{tab:yearsofobservations-mostpromisingstars}.

In respect of the black hole no-hair theorem, we conclude that with S2 orbit it is not possible to test the theorem even with 40 years of daily measurements using GRAVITY's resolution limits on S2 orbit in Scenario IV in Table \ref{tab:scenarios-summary}.
In order to test the no-hair theorem with GRAVITY's best resolution, we need a closer star. It is expected to find stars with orbital periods of 1-2 years by the next generation large telescopes \citep{Graham:2019mis}. In our simulations we use a stellar orbit around the Galactic Center that is one fifth the size of S2 orbit.  With such a star, we can start to measure the quadrupole moment and test the black hole no-hair theorem with 20 years of weekly observations as shown in Figure \ref{fig:onefifthS2orbitQ2test} for Scenario V. It is necessary to understand the distribution of visible and dark matter outside the black hole to better constrain the no-hair theorem; however, without such knowledge, we can treat the quadrupole moment as an independent parameter on the black hole spin and a term that combines the quadrupole moment of both the black hole and the external sources in the vicinity of Sgr A*, and see how well we can measure it as shown in Figure \ref{fig:onefifthS2orbitQ2test}.  

Several other scenarios can be further studied based on our investigation. The epochs of the observations are equally spaced in our simulations. If these measurements are rearranged such that they are more frequently made when the star is close to the periapsis of its orbit around the black hole than the apoapsis, the parameter measurement uncertainties in the model are expected to reduce. Another factor to consider is that the telescope measurement resolution can be changed in future scenarios. In this work it is chosen to be the limits of GRAVITY telescope for all the fake observations we present. However, if the resolution can be further improved, we can have better observation strategies using less observation time to test the black hole no-hair theorem. In addition, if several more closer stellar orbits are found then we can use them to jointly constrain the black hole properties.

Another direction to explore is to use the radio images of supermassive black holes. The Event Horizon Telescope measurements are complementary to stellar proper motions and therefore could break some degeneracies and make constraints on the black hole nature of the central remnant easier \citep{Psaltis:2018xkc,Akiyama:2019cqa,2020AAS...23536915D,Reynolds:2019uxi,Dokuchaev:2019bbf,Bambi:2019tjh, Psaltis:2020lvx}.

In Section \ref{sec:ambientperturbers} we briefly discuss the perturbing effects from a stellar cluster and an IMBH on our interpretation of the orbits. However, existing observations do not provide enough information to constrain the ambient density of perturbers. Dynamical processes have long been expected to produce a high density of nearby massive objects, as yet inaccessible to direct electromagnetic observation
\citep{2006ApJ...649...91F,2010ApJ...708L..42P,2009ApJ...697.1861A,2010ApJ...718..739M,2010PhRvD..81f2002M,2010RvMP...82.3121G,2014ApJ...780..148A}.  
This dark density is most likely to be constrained indirectly, via its gravitational effects (e.g.,
\citep{2014ApJ...780..148A}, though \citep{2013PhRvL.110v1102B}).    
Anisotropies in the ambient density can partially mimic the effects of modified theories of gravity; for example, a
quadrupolar gravitational perturbation could be sourced by the black hole or an external cluster density.

\begin{acknowledgements}
We thank Phil Chang for his extensive help with the Fortran orbit evolution solver which crucially speeds up the code, and Clifford Will for useful discussions on the project. The project was supported by the STFC grant ST/T000147/1. We are grateful for the computational resources that were supported by NSF-0923409 and provided by the Leonard E Parker Center for Gravitation, Cosmology and Astrophysics at University of Wisconsin-Milwaukee, where Hong conducted most of the calculations for this work as a graduate student.
\end{acknowledgements}

\onecolumngrid
\appendix
\section{Equations of motion}
\label{ap:EOMs}

In this appendix, we point out that different properties of an ensemble of stellar orbits probe different physics.  For
example, the orbit location probes different parts of the potential: distant
orbits preferentially probe an external potential while nearby orbits probe the black hole.   Similarly, different
symmetry-breaking effects only occur from certain physical processes; for example, spherically symmetric potentials cannot cause
the orbital plane to precess, while quadrupolar Newtonian potentials and frame dragging cause an ensemble of orbits to
evolve in distinctly different ways.  
By isolating these symmetries and their impact on observations, we can easily model how a collection of measurements of
several stellar orbits can best constrain properties of the Galactic Center environment.

In the text, we adopted simple approximations to general relativity at low post-Newtonian order, neglecting many common
factors like the mass ratio.    Because orbital perturbations we hope to identify are small, influences from small
factors like mass ratio ($\simeq 10^{-6}$) can be of similar order to the minute effects we seek to identify at
targeted separations.    For this reason, in this section we carefully review relevant post-Newtonian expressions,
targeting typical separations (i.e., 10 year orbits) and post-Newtonian accuracy ideally comparable to the targeted
astrometric resolution of $\mu$as per year at 8 kpc 
(i.e.,  $\simeq 0.26$ Myear, or $\Delta v/c\simeq 10^{-7}$).

Post-newtonian theory for binary and N-body motion is well-developed;  see  \cite{book-merritt-GalacticCenter} for a
review in the context of stellar orbits around supermassive black holes; \cite{ACST} for a
discussion of orbit-averaged spin-precession; and \cite{lrr-Blanchet-PN},
\cite{book-Will-TestingGR} for  technically sophisticated and highly detailed discussions in general and for binary
motion, specifically. \\

\subsection{Post-Newtonian Two-body equations of motion}

Working to $v^2$ (1PN) beyond Newtonian order in velocity and leading-order in spin-orbit coupling, the post-Newtonian Lagrangian for two-body motion has the form \cite{book-merritt-GalacticCenter}
\begin{eqnarray}
{\cal L} = \eta M \left[\frac{1}2 v^2 + \frac{GM}{r} + \frac{1}{8}(1-3 \eta) v^4 + \frac{GM}{2r}(3+\eta)v^2+\eta
  \dot{r}^2 - \frac{GM}{r} \right]
 + {\cal L}_{spin} + {\cal L}_{quad},
\end{eqnarray}
using units with $c=1$ for simplicity. Here ${\cal L}_{spin}$ and ${\cal L}_{quad}$ terms are due to the black hole spin and the quadrupole moment.
The Lagrangian corresponds to the Hamiltonian \cite{2006PhRvD..74j4005B}
\begin{align}
&H =\mu[ H_N+H_{1PN}+ H_{SO}] \\
&H_N = \frac{p^2}{2}  - \frac{M}{r} \\
&H_{1PN} = \frac{1}{8}(3\eta-1)p^4 - \frac{1}{2}[(3+\eta)p^2+\eta p_r^2]\frac{M}{r}+ \frac{M^2}{2r^2} \\
&H_{SO} = 2 \frac{{\bf L}_N/\mu\cdot {\bf J} }{r^3}.
\end{align}
These approximations, plus the limit $\eta\rightarrow 0$, reproduce the equations of motion adopted in the text.  
These Hamiltonian expressions also enable straightforward derivation of the orbit-averaged precession equations.  As a
concrete example, the contribution of black hole spin to the orbit-averaged precession equations for $L_N,A_N$ follow
from the Lie algebra 
\begin{align}
(\partial_t L_a)_{SO} &= \poissonbracket{L_a}{H_{SO}}  = \frac{2 \epsilon_{abc}J_b L_c}{r^3} \\
(\partial_t A_a)_{SO} &= \poissonbracket{(p\times L - M \hat{r})_a}{H_{SO}} 
 \nonumber\\&=  \poissonbracket{(p\times L - M \hat{r})_a}{\frac{1}{r^3}} (2\vec{J}\cdot \vec{L})
 +  \poissonbracket{(p\times L - M \hat{r})_a}{L_d} 2J_d/r^3 
\nonumber\\&
= -3\frac{\epsilon_{abc} r_b L_c}{r^5} (2\vec{J}\cdot \vec{L}) +  \epsilon_{abc}\frac{2  J_b }{r^3} A_c,
\end{align}
using $\poissonbracket{L_a}{V_b} = \epsilon_{abc}V_c$ for any vector $V$ rotating with $L$ (here,
$\vec{L},\vec{p},\vec{r}$).   Both orbit averages can be performed trivially, substituting $\vec{r}=p(\hat{x} \cos
\theta+\hat{y}\sin\theta)/(1+e \cos \theta)$ and $dt=d\theta L/r^2$ for the special case $\vec{A}=e\hat{x}$; we find 
\begin{align}
\left<r^{-3}\right> & = \frac{2\pi}{P} \frac{M}{p^3} \\
\left<r \cos \theta r^{-5}\right> &
= \frac{2\pi}{P} \frac{e M}{p^3}. 
\end{align}
Critically, the second term does   \emph{not} orbit-average to zero.  We therefore find 
\begin{eqnarray}
\left< (\partial_t A)_{\rm SO} \right> = \frac{2M}{p^3} \left[ \vec{J} - 3(\vec{J}\cdot \hat{L}) \hat{L}   \right] \times A.
\end{eqnarray} 

Are these approximations adequate?   First and foremost, as emphasized in the text, most post-Newtonian and mass ratio
effects do not break symmetry in a way that can be confused with the influence of precession: even if they \emph{did}
matter quantitatively, they wouldn't matter qualitatively.  
Second, for a \emph{single} star, the back-reaction of the star on the BH's orbit is small at typical high mass ratio ($\eta
\simeq 10^{-6}$);  the leading-order effect is purely Newtonian,  corresponding to orbits around the center of mass; and
higher-order PN effects are suppressed by $O(v^2)\simeq M/r\simeq 10^{2}-10^3$.   For a single star, the finite mass
ratio is a minute perturbation at separations where precession can be measured astrometrically; see  \ref{fig:PrecessionRates}.

As emphasized in the text, however, this modification does not break symmetry and therefore does not significantly
influence the quantitative accuracy to which precession-induced modulations can be measured. 

\subsection{Post-Newtonian N-body equations of motion}
When many bodies are included, we must carefully account for the often significant perturbations from neighboring
stars, as well as the collectively weakly significant reaction of the black hole to the ambient stellar potential.


Finally, the BH spin will precess  to conserve total angular momentum as the stars precess
\cite{book-merritt-GalacticCenter} due to Lens-Thirring effects, as well due to the ambient gravitational potential \cite{2014arXiv1404.5160H}. 
As the spin precesses, the leading-order spin-orbit precession will be modulated, an effect that can be comparable to
 quadrupolar precession effects from the central supermassive black hole.  


\section{Fisher matrix for Newtonian orbits}
To constrain properties of the Galactic Center, we must first identify the Newtonian orbit.  In this section we review
how to calculate the Fisher matrix for Newtonian orbital parameters using our toy-model likelihood equation for special cases and in relative generality.

\subsection{Fisher matrix for Keplerian orbits}
\label{ap:fisher-keplerian}
In the discussion above, we adopted as coordinates the initial velocity and position.  This choice of coordinates is
particularly compatible with our equations of motion and subsequent analytic calculations (e.g., including non-Newtonian
perturbations).  
While straightforward for brute-force calculations, the above approach is rarely analytically tractable.  
Alternatively, the perturbed orbit $\Delta r$ can be reduced to (a) changes of $a,e$ and the Newtonian orbital phase $\Phi_0$ and (b) changes in the
orientation of the orbit.  Using the chain rule, we can build up the total perturbation as an additive
contributions from both factors, each individually simple and particularly tractable in suitable coordinates.

Specifically, using as coordinates the orientation of the orbital frame (3 parameters) as well as $a,e,\Phi_0$ (3
parameters),  we can express
\begin{align}
\Delta \vec{r}(t)&=\vec{C}_{a}(t)\Delta a +\vec{ C}_{e}\Delta e+\vec{C}_{\Phi}\Delta\Phi_0 + (-i {\cal L}_\beta \vec{r})\Delta \Theta^\beta,
\end{align}
where $C_{\alpha,X}$ for $\alpha=x,y,z$ are the Cartesian components of the vectors $\vec{C}_{X}$ and where $\Delta \Theta^\beta$ is a small (constant) rotation vector and ${\cal L}_\alpha$ are the generators of rotations.
As a concrete example, for circular orbits $\vec{r}=a[\cos (\Omega_N T) \hat{x}+\sin (\Omega_N T)\hat{y}]$, with T as the observation time and $\Omega_N$ the rotation rate of the star
\begin{align}
\vec{C}_{a}&= \hat{r} + \frac{\partial \Omega_N}{\partial a}T a \hat{v} \\
\vec{C}_{\Phi}&= a \hat{v} \\
\vec{C}_{e} &= 0.5a \{[-3+\cos (2\Omega_N T) ]\hat x + \sin(2\Omega_N T) \hat y\}\\
-i {\cal L}_x \vec{r}&=[\hat{y}\hat{z}-\hat{z}\hat{y}]_{ab}\vec{r}_b=-a \hat{z}(\hat{r}\cdot \hat{y}) \\
-i {\cal L}_y \vec{r}&=[-\hat{x}\hat{z}+\hat{z}\hat{x}]_{ab}\vec{r}_b =a\hat{z}(\hat{r}\cdot \hat{x})
\\
-i {\cal L}_z \vec{r}&=[\hat{x}\hat{y}-\hat{y}\hat{x}]_{ab}\vec{r}_b =a\hat{x}(\hat{r}\cdot \hat{y})-a\hat{y}(\hat{r}\cdot \hat{x})= -a\hat v,
\end{align}
and rotations around $z$ are degenerate with the change in orbital reference phase $\Phi_0$. 

In terms of these coordinates, the Fisher matrix for the idealized measurements in
Eq.~\eqref{eq:MeasurementModel:IdealPosition} can be expressed in the particularly analytically tractable form
\begin{eqnarray}
\Gamma_{\alpha \beta}&=\frac{N}{\sigma_r^2}
\begin{bmatrix}
\int \frac{dt}{T} \sum_b C_{b,a} C_{b,a} &  \int \frac{dt}{T}  \sum_b C_{b,a}C_{b, e} &  \int \frac{dt}{T}  \sum_b C_{b,a}C_{b, \Phi}    & \int \frac{dt}{T} \sum_b C_{b,a} [-i {\cal L}_\beta \vec{r}]_b \\
 \int \frac{dt}{T}  \sum_b C_{b,e}C_{b, a} & \int \frac{dt}{T}  \sum_b C_{b,e}C_{b, e} &  \int \frac{dt}{T}  \sum_b C_{b,e}C_{b, \Phi}  & \int \frac{dt}{T} \sum_b C_{b,e} [-i {\cal L}_\beta \vec{r}]_b \\
 \int \frac{dt}{T}  \sum_b C_{b,\Phi}C_{b, a} & \int \frac{dt}{T}  \sum_b C_{b,e}C_{b, \Phi} &  \int \frac{dt}{T}  \sum_b C_{b,\Phi}C_{b, \Phi}  & \int \frac{dt}{T} \sum_b C_{b,e} [-i {\cal L}_\beta \vec{r}]_b \\
\int \frac{dt}{T} \sum_b C_{b,a} [-i {\cal L}_\beta \vec{r}]_b & \int \frac{dt}{T} \sum_b C_{b,e} [-i {\cal L}_\beta
  \vec{r}]_b & \int \frac{dt}{T} \sum_b C_{b,\Phi} [-i {\cal L}_\beta \vec{r}]_b  &  \int \frac{dt}{T} [{\cal L}_\alpha \vec{r}] \cdot [{\cal L}_\beta \vec{r}] 
\end{bmatrix}.
\end{eqnarray}
We confirm this representation reproduces the results provided above. Being analytically tractable even for eccentric orbits, this general form is particularly  well-suited to marginalization via
Eq.~\eqref{eq:Fisher:Marginalize}.   

For circular orbits, the expressions involved can be approximately evaluated, using the following rules 
\begin{align}
\left<\hat{r}_a\hat{r}_b\right> =\frac{1}{2}[ \delta_{ab}-\hat{L}_a\hat{L_b}] \\
\left<\hat{v}_a\hat{v}_b\right> =\frac{1}{2}[ \delta_{ab}-\hat{L}_a\hat{L_b}] \\
\left<\hat{r}_a\hat{v}_b\right> = 0, 
\end{align}
and by applying these rules, we find the expressions for the Fisher matrix components:
\begin{align}
\label{eq:FisherMatrixRadiusComponentforKeplerianOrbits}
\Gamma_{aa} &=\frac{N}{\sigma_r^2} \int \frac{dt}{T} \sum_b C_{b,a} C_{b,a} 
 =\frac{N}{\sigma_r^2} \int \frac{dt}{T} (1+t^2 a^2(\partial \Omega_N/\partial a)^2)  \\
\Gamma_{\Phi\Phi}&= \frac{N a^2}{\sigma_r^2}\\
\Gamma_{ee}&=\frac{5N a^2}{2\sigma_r^2} \\
\Gamma_{ae}&=0 \\
\Gamma_{a\Phi}&=\Gamma_{a\Theta_z}=\frac{N}{\sigma_r^2} \int \frac{dt}{T} t a^2(\partial \Omega_N/\partial a)   \\
\Gamma_{e\Phi}&=0 \\
\Gamma_{\Theta_x a} &=\Gamma_{\Theta_y a} =\Gamma_{\Theta_x e} = \Gamma_{\Theta_y e} =\Gamma_{\Theta_x \Phi} = \Gamma_{\Theta_y \Phi}=0\\
\Gamma_{\Theta_x\Theta_x}&=\Gamma_{\Theta_y\Theta_y} =\frac{Na^2}{2\sigma_r^2}\\
\Gamma_{\Theta_y\Theta_y} &=\frac{Na^2}{\sigma_r^2}\\
\Gamma_{\Phi\Theta_z} &=-\frac{Na^2}{\sigma_r^2}.
\end{align}
The terms in this circular-orbit Fisher matrix have qualitatively different behavior.  On the one hand, changes in the orbital period
($a$) lead to significant, increasing dephasing across multiple orbits; as a result, the orbital radius can be measured
with high accuracy, increasing rapidly as the measurement interval increases [$\Gamma_{aa} \propto (\omega T)^2 N
(a/\sigma_r)^2$].    By contrast, all other changes in a circular orbit are \emph{geometrical}, producing
\emph{small or variable} separations.  While our ability to measure these parameters also increases with the number of
measurements ($\propto N \propto T$), the accuracy to which  these parameters can be measured  is significantly
smaller. 
Finally, the circular-orbit Fisher matrix decomposes trivially into diagonal terms (almost all) plus one $2\times2$ block
($\ln a,\Phi$); this
nearly-degenerate $2\times2$ block can be trivially diagonalized
\begin{align}
\Gamma_{ab}=
\frac{Na^2}{\sigma_r^2}\begin{bmatrix}
1+\frac{T^2}{3} a(\partial_a\Omega)^2 & \frac{T}{2} a(\partial_a\Omega) \\
 \frac{T}{2} a(\partial_a\Omega) &a^2
\end{bmatrix}
=\frac{Na^2}{\sigma_r^2}\begin{bmatrix}
1+\frac{3}{4}\Phi_{orb}^2 & -\frac{9}{8}\Phi_{orb} \\
-\frac{9}{8}\Phi_{orb} & \frac{9}{4}\Phi_{\rm orb}^2
\end{bmatrix}
\end{align}
using $Ta \partial_a\Omega_N =-3\Phi_{orb}/2$ for $\Phi_{orb}=\Omega_N t$ the orbital phase.  The relative significance
of the two terms depends on how many orbital cycles have occurred.  


\subsection{Unknown black hole mass}
Adding additional parameters, like the black hole mass, is straightforward:
\begin{eqnarray}
\Delta\vec{r}(t) &= \sum_A \vec{C}_{\lambda} \Delta \lambda.
\end{eqnarray}
For circular orbits, the effect of a perturbed black hole mass is \emph{very similar} to a perturbed orbital separation,
producing a significant dephasing with time without any  (small) change in position:
\begin{eqnarray}
\vec{C}_{M}&= \frac{\partial \Omega_N}{\partial M}t a \hat{v}.
\end{eqnarray}
Because the Newtonian orbital period only depends on $\sqrt{M/a^3}$, these two parameters are nearly degenerate in the
Fisher matrix: we can only measure one combination (the orbital period!) reliably.
Marginalizing out the unknown orbital radius $a$, we find
the Fisher matrix for black hole parameters does \emph{not} depend as sensitively on the stellar mass.  For circular
orbits specifically, all parameters except $M,a,\Phi$ separate, allowing us to marginalize only a 3-dimensional matrix
\begin{eqnarray}
\Gamma_{MM}&= \frac{Na^2 t^2}{3\sigma_r^2}\left(\frac{\partial \Omega_N}{\partial M} \right)^2.
\end{eqnarray}

\subsection{Unknown black hole spin}
\label{ap:fisher-spin}
The black hole spin enters via $\vec{\Omega}$ in a particularly simple way at leading order:
$\partial\Omega^\alpha/\partial J_\beta = \delta^\alpha_\beta Z_J$. For example, the Fisher matrix over $J$ components
has the form
\begin{align}
\Gamma_{\alpha \beta} &= \frac{N}{\sigma_r^2}\int \frac{dt}{T} \frac{\partial \Omega^a}{\partial J^\alpha}\frac{\partial \Omega^b}{\partial J^\beta}
\left< t^2 [{\cal L}_a r_o]\cdot [{\cal L}_b r_o] \right> \nonumber\\ 
&\simeq \frac{N}{\sigma_r^2} \frac{\partial \Omega^a}{\partial J^\alpha} \frac{\partial \Omega^b}{\partial J^\beta}
\frac{T^2}{3} \text{Tr}[{\cal L}_a I {\cal L}_b^T] \nonumber\\
&\textcolor{black}{\simeq \frac{N Z_J^2T^2}{3\sigma_r^2} \int_0^P\frac{dt}{P} \text{Tr}[{\cal L}_a I {\cal L}_b^T] } \nonumber\\
&\textcolor{black}{=\frac{N Z_J^2 T^2 [a(1-e^2)]^4}{3\sigma_r^2 P L} \text{Tr}[{\cal L}_a (A_1\hat x\hat x + (A_2-A1)\hat y\hat y){\cal L}_b^T]},
\label{eq:Intermediate:FisherVersusJ}
\end{align} 
where 
\begin{eqnarray}
A_1\equiv\int_0^{2\pi } d\theta\ \frac{\cos^2\theta }{(1+e\cos\theta)^4}
=\frac{(1+4 e^2)\pi}{(1-e^2)^{7/2}}
\end{eqnarray} 
and 
\begin{eqnarray}
A_2\equiv\int_0^{2\pi} d\theta\ \frac{1}{(1+e\cos\theta)^4}
=\frac{(2+3 e^2)\pi}{(1-e^2)^{7/2}}.
\end{eqnarray}
Both integrals can be performed analytically when $T/P$ is an integer; in this special case we find
\begin{align}
A_2 &\simeq \frac{ T}{P} \int_0^{2\pi} d\theta \frac{1}{(1+e \cos \theta)^4} 
  = 2\pi T /P   \frac{(1+ \frac{3}{2}e^2)}{(1-e^2)^{7/2}} \\
A_1 &= \pi T /P  \frac{1+4 e^2}{(1-e^2)^{7/2}}.
\end{align}
Using the explicit form of the generators ${\cal L}$ in this frame, we find the trace 
\begin{align}
\text{Tr}[{\cal L}_a (A_1\hat x\hat x + (A_2-A1)\hat y\hat y){\cal L}_b^T] = \left[ \begin{array}{ccc} A_2-A_1& 0 & 0 \\ 0& A_1 & 0 \\ 0 & 0 & A_2 \end{array} \right].
\end{align}
The $\Gamma_{J_aJ_b}$ components are then a coefficient times a matrix, and the other matrix components are expressed as following
\begin{align}
\label{eq:FisherMatrixSpinComponentforGenericOrbits}
\Gamma_{J_aJ_b} &= \frac{2 N T^2(1-e^2)^{1/2}}{3\pi\sigma_r^2 a^4} \left[ \begin{array}{ccc} A_2-A_1& 0 & 0 \\ 0& A_1 & 0 \\ 0 & 0 & A_2 \end{array} \right]\\
\Gamma_{a J_x} &= \frac{4\pi NT^2}{\sigma_r^2 a^5} Tr[(J_x{\cal L}_x + J_y{\cal L_y} + J_z{\cal L}_z )[\hat x\hat x + \hat y\hat y]{\cal L}_x^T]\\
\Gamma_{a J_y} &= \frac{4\pi NT^2}{\sigma_r^2 a^5} Tr[(J_x{\cal L}_x + J_y{\cal L_y} + J_z{\cal L}_z )[\hat x\hat x + \hat y\hat y]{\cal L}_y^T]\\
\Gamma_{a J_z} &= \frac{-8\pi NT^2}{\sigma_r^2 a^5} Tr[(J_x{\cal L}_x + J_y{\cal L_y} + J_z{\cal L}_z )[\hat x\hat x + \hat y\hat y]{\cal L}_z^T]\\
\Gamma_{e J_x} &=\Gamma_{e J_y} =0\\
\Gamma_{e J_z} &= \frac{NT}{\sigma_r^2 a}\\
\Gamma_{\Phi_0 J_x} &=\Gamma_{\Phi_0 J_y} =\Gamma_{\Phi_0 J_z}=0\\
\Gamma_{\Theta_x J_x} &= \frac{NT}{2\pi\sigma_r^2 a}Tr[{\cal L}_x[A_1\hat x\hat x +(A_2-A_1)\hat y\hat y]{\cal L}_x^T]\\
\Gamma_{\Theta_y J_y} &= \frac{NT}{2\pi\sigma_r^2 a}Tr[{\cal L}_y[A_1\hat x\hat x +(A_2-A_1)\hat y\hat y]{\cal L}_y^T]\\
\Gamma_{\Theta_z J_z} &= \frac{-NT}{2\pi\sigma_r^2 a}Tr[{\cal L}_z[A_1\hat x\hat x +(A_2-A_1)\hat y\hat y]{\cal L}_z^T].
\end{align}

\section{Likelihood and MCMC }
\label{sec:toyincartesian}
\subsection{Bayesian formalism}
To separate issues pertaining to measurement from physics from dynamics, we describe results using three measurement
scenarios: (a) an idealized measurement model, where the position or velocity of each star can be measured  at known times, as if
via an array of local observers surrounding the black hole; (b) a plausible model, where only the radial velocity and
transverse angle can be measured, on known null rays; and (c) a model for pulsar timing.

Specifically, our first measurement model assumes  each star's position $\vec{r}_\alpha$ is measured to be $\vec{x}_{\alpha,k}$ on times $t_k$ with measurement error
$\sigma_r$.   We will henceforth use Greek subscripts $\alpha$ to index stars \emph{or} parameters; small roman subscripts like $k$ to index
  measurements; and large roman symbols to denote vector components.   Since local measurements are performed, the distance to the black hole (and astrometry) do not enter into
the analysis.    For this model, the probability distribution of the data is
\begin{eqnarray}
\label{eq:MeasurementModel:IdealPosition}
p(D|\lambda) = (2\pi \sigma_r^2)^{-3N/2} \exp - \sum_{\alpha,k} \frac{(\vec{r}_\alpha(t_k|\lambda) - \vec{x}_k)^2}{2\sigma_r^2}.
\end{eqnarray}
Because of its simplicity, we will use this analytically trivial model  when illustrating how physics break the degeneracy.

A more realistic measurement model accounts for the unknown distance to the Galactic Center; the unknown mass of the
Galactic Center black hole; and the fact that only projected sky positions $\vec{\theta}_k$ and radial velocities $v_{r,k}$ can be measured.   For
this model, the probability distribution of the data are
\begin{align}
p(D|\lambda) &=
 (2\pi \sigma_\theta^2)^{-2N/2} \exp - \sum_{\alpha,k} \frac{(P_\perp\vec{r}_\alpha(t_k|\lambda) -  \vec{\theta}_kR)^2}{2\sigma_\theta^2}
\nonumber \\
&\times  (2\pi \sigma_v^2)^{-N/2} \exp - \sum_{\alpha,k} \frac{(\hat{N}\cdot \partial_t\vec{r}_\alpha(t_k|\lambda) -  v_N)^2}{2\sigma_v^2},
\end{align}
combined with a prior for $R$, the distance to the Galactic Center.   
A more realistic model still accounts for light propagation time across the stellar orbit \cite{Fupeng2015}; light bending near the black hole 
 note we are in harmonic coordinates; higher order terms in the doppler equation \cite{2007ApJ...654L..83Z,2006ApJ...639L..21Z}

Finally, the orbit of a pulsar around a black hole can be reconstructed by timing.   Pulsar timing
corresponds to fitting a model to pulse arrival times, to insure they arrive in regular intervals in the source frame.
Roughly speaking, the model corresponds to fitting the proper time of the pulsar's orbit, which can be measured to some
accuracy. 

\subsection{Fisher matrix}
To illustrate the mechanics of a Fisher matrix calculation,  we employ the idealized measurement model of
Eq.~\eqref{eq:MeasurementModel:IdealPosition} in the special case that the observed data is exactly as predicted by some
set of model parameters $\lambda'$ [i.e., $\vec{x}_k  = \vec{r}(t_k|\lambda')$].  Using a first-order Taylor series
expansion  $\vec{r}(t_k|\lambda)-\vec{r}(t_k|\lambda')\simeq \delta \lambda^b \partial \vec{r}/\partial \lambda_b$ for
the position versus parameters $\lambda$, we find the conditional probability of the data given $\lambda$ can be
approximated by
\begin{align}
\ln p(D|\lambda)  &= \text{const} - \frac{1}{2} \Gamma_{ab} \delta \lambda_a \delta \lambda_b \\
\label{eq:numericalfisher}
\Gamma_{ab} &= \sum_{\alpha,k}\frac{1}{\sigma_r^2} \frac{\partial \vec{r}_{\alpha}}{\partial\lambda_a}\frac{\partial \vec{r}_{\alpha}}{\partial\lambda_b}.
\end{align}
This expression applies in general, no matter how the solution $r(t)$ is solved or approximated.
By using an approximate analytic solution, the orbit-averaged secular solution in Eq.~\eqref{eq:OrbitAveragedSolution:SecularRotation}, we can estimate the accuracy to which parameters can be
measured using a simple orbit average over a Newtonian solution.  For example, for parameters $\lambda$ which do not
appear in the unperturbed Newtonian solution, like the black hole spin $J$ or external potential, the Fisher matrix takes the form
\begin{align}
\Gamma_{ab} &=\sum_{k} \frac{t_k^2}{\sigma_r^2}\frac{\partial\Omega^A}{\partial\lambda_a}  \frac{\partial\Omega^B}{\partial\lambda_b} (-i {\cal L}_A \vec{r}_o(t_k))^C(-i {\cal L}_B \vec{r}_o(t_k))^C.
\end{align}
In fact, as a first approximation, these components of Fisher matrix can be approximated using the orbit's moment of
inertia $I_{ab,N} = \left<r_{o,a} r_{o,b}\right>$:
\begin{eqnarray}
\Gamma_{ab} &\simeq \frac{t^3}{3 N}\frac{\partial\Omega^A}{\partial\lambda_a}
\frac{\partial\Omega^B}{\partial\lambda_b}
\text{Tr}[(-i {\cal L}_A) I (-i {\cal L}_B)^T].
\end{eqnarray}
Having estimated the Fisher matrix and hence approximated $p(\{d\}|\lambda)$ by a Gaussian, we can further construct
marginalized distributions for $\lambda_A$ in $\lambda=(\lambda_A,\lambda_a)$ by integrating out the variables
$\lambda_a$.   

\subsection{Toy model: tests in \texorpdfstring{$\vec r$} using MCMC}
\label{ap:toy}

\begin{figure}[ht]
\centering
\includegraphics[trim=2.3cm 0.cm 3.6cm 1.6cm, clip=true, width=0.495\columnwidth]{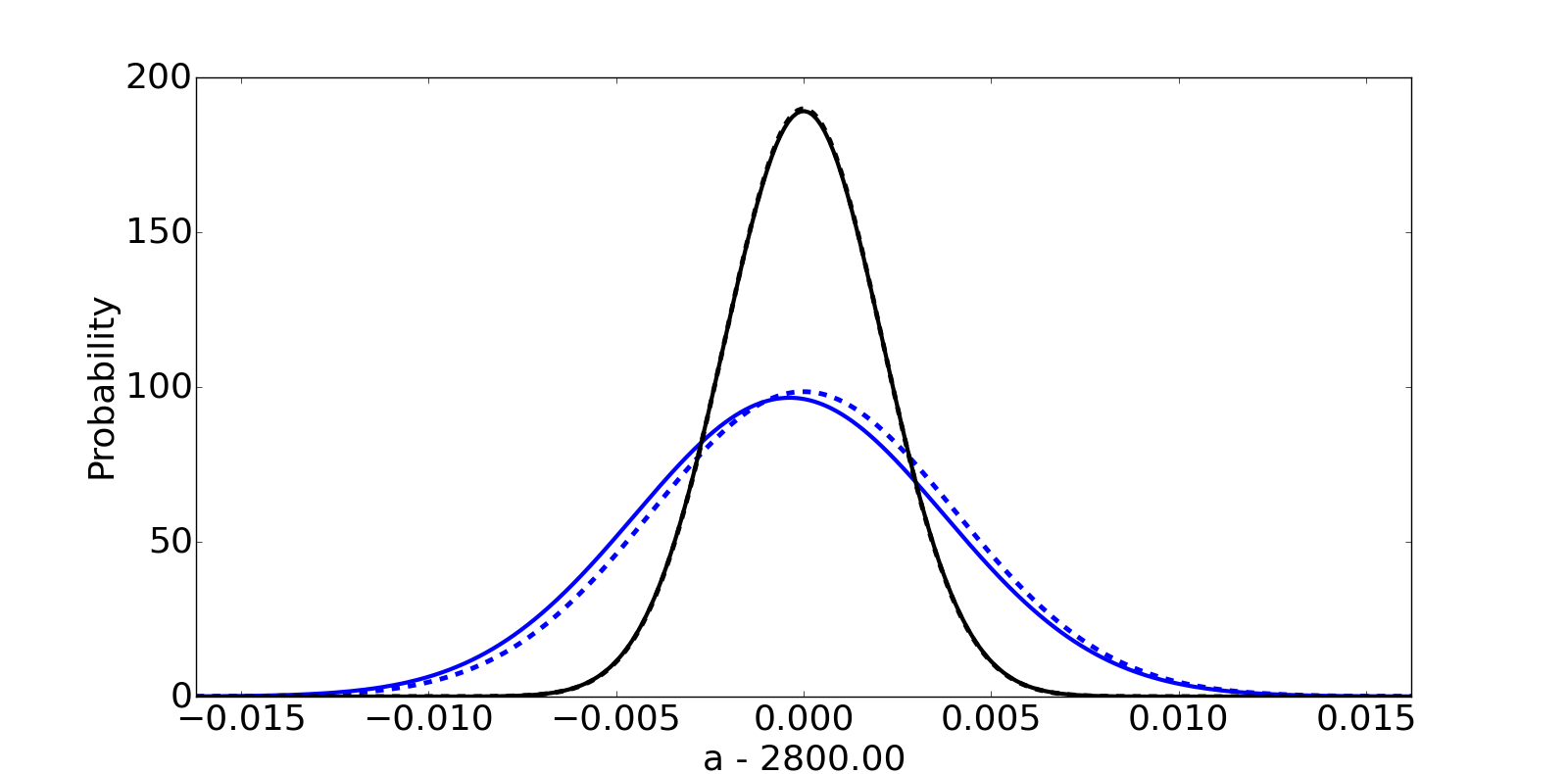}
\includegraphics[trim=2.8cm 0.cm 3.1cm 1.5cm, clip=true, width=0.495\columnwidth]{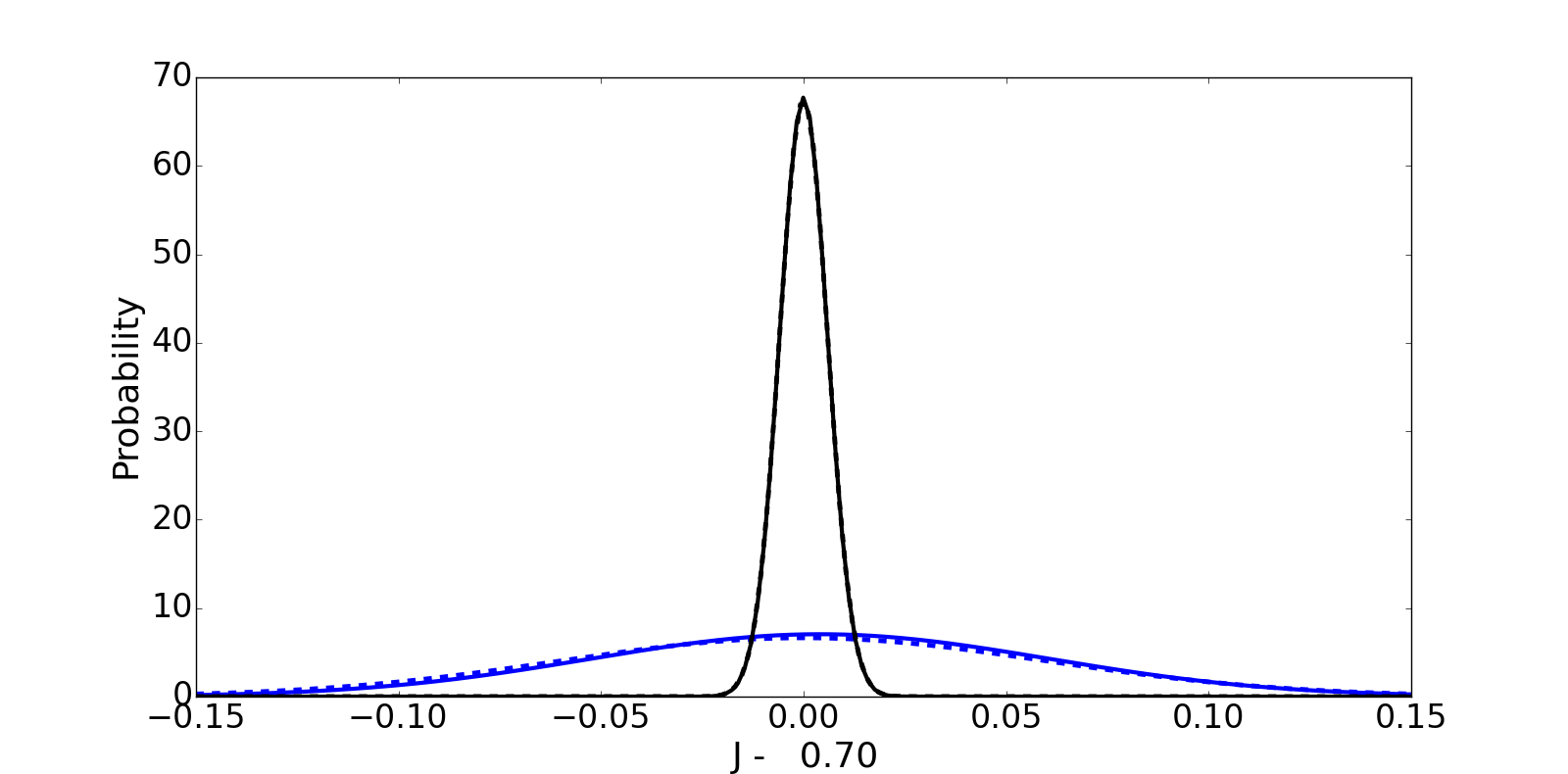}
\caption{\label{fig:CalibrateResults:CircularOrbit}\textbf{Measuring properties of an orbit and a BH for idealized measurement model.} 
\emph{Left panel}: Demonstration of how accurately the radius of a Newtonian circular orbits can be measured by our MCMC code, assuming the only unknown parameter is the
orbital radius (black) and assuming no parameters are known (blue).  For comparison, the dotted curves show the results
from our Fisher matrix calculations.
\emph{Right panel}: Demonstration of how accurately the black hole spin $J$ can be measured, assuming the only unknown
parameter is the black hole spin magnitude $\chi=J/M^2$ (black) and assuming both the orbit and black hole spin vector are unknown (blue). The dotted curves show the result from Fisher matrix calculations.
\label{fig:toymodel}
}
\end{figure}

We show that MCMC agree with both the numerical and the analytic Fisher matrices via toy models: As a concrete example, in the Cartesian coordinates with its origin at the black hole center and $\{x_i, y_i, z_i\}$ as the observables, we model a Newtonian circular orbit with parameters $\{a, \Phi_0, \beta, \gamma, \psi\}$ and measure its semimajor axis or radius in two cases shown in the left panel of Figure \ref{fig:toymodel}, as well as an elliptical orbit with parameters $\{a, e, \Phi_0, \beta, \gamma, \psi, J_x, J_y, J_z\}$ and measure its spin magnitude in two cases shown in the right panel of Figure \ref{fig:toymodel}. For the measurement of the radius (denoted with symbol $a$ as it is semimajor axis with $e=0$) of a circular orbit, the two cases are treating only the semimajor axis as uncertain as shown in the black solid line and treating all orbital parameters as uncertain as shown in the blue solid line. The dashed lines show the measurement uncertainty from the Fisher matrix method where both the numerical Fisher matrix in Eq.~\eqref{eq:numericalfisher} and the analytic Fisher matrix component for the radius in Eq.~\eqref{eq:FisherMatrixRadiusComponentforKeplerianOrbits} give the same value, with $a=2800\ M,\ N=700,\ \sigma_r=1.0\ M,\ T=100\ \text{week},\ \text{and}\ \Delta t=1\ \text{day}$. Comparing the corresponding solid and the dashed lines for the two cases respectively, we can see that MCMC agree well with Fisher matrix for the measurement uncertainties in the radius of the orbits. Similar conclusion can be drawn for the measurement of the magnitude of black spin. Note that the numerical Fisher matrix in Eq.~\eqref{eq:numericalfisher} and the analytic Fisher matrix in Eq.~\eqref{eq:FisherMatrixSpinComponentforGenericOrbits} are used and they are also the same. They are evaluated using the same initial parameters as the left panel, except that $\sigma_r=0.1\ M$ and $e=0.01$ and evolved according to Eq.~\eqref{eq:EOM:SingleBody}. 
\newline
\bibliography{gco}

\end{document}